\documentclass[sigconf,screen]{acmart}
\usepackage[strings]{underscore}  
\acmSubmissionID{papers_402}
\usepackage{soul}
\usepackage{framed}

\usepackage{caption}
\usepackage{subcaption}

\newcommand{\argmax}[0]{\text{argmax}}
\newcommand{\p}[0]{\textsf{P}}
\newcommand{\E}[0]{\textbf{E}}

\definecolor{kc}{rgb}{1,0.9,0.9}

\definecolor{jrk}{rgb}{0.9,0.9,1.0}

\definecolor{tml}{rgb}{0.7,0.9,0.7}

\definecolor{jbt}{rgb}{0.9,0.9,0.9}

\title{Acting as Inverse Inverse Planning}
\author{Kartik Chandra}\email{kach@csail.mit.edu}
\affiliation{\institution{MIT CSAIL}\city{Cambridge}\country{USA}}
\author{Tzu-Mao Li}\email{tzli@ucsd.edu}
\affiliation{\institution{UC San Diego}\city{San Diego}\country{USA}}
\author{Josh Tenenbaum}\email{jbt@mit.edu}
\affiliation{\institution{MIT BCS and CSAIL}\city{Cambridge}\country{USA}}
\author{Jonathan Ragan-Kelley}\email{jrk@mit.edu}
\affiliation{\institution{MIT CSAIL}\city{Cambridge}\country{USA}}

\ccsdesc[500]{Computing methodologies~Procedural animation}
\ccsdesc[500]{Computing methodologies~Cognitive science}
\ccsdesc[500]{Computing methodologies~Theory of mind}

\keywords{storytelling, animation, inverse planning, Bayesian inference}

\copyrightyear{2023} 
\acmYear{2023} 
\setcopyright{rightsretained} 
\acmConference[SIGGRAPH '23 Conference Proceedings]{Special Interest Group on Computer Graphics and Interactive Techniques Conference Conference Proceedings}{August 6--10, 2023}{Los Angeles, CA, USA}
\acmBooktitle{Special Interest Group on Computer Graphics and Interactive Techniques Conference Conference Proceedings (SIGGRAPH '23 Conference Proceedings), August 6--10, 2023, Los Angeles, CA, USA}
\acmDOI{10.1145/3588432.3591510}
\acmISBN{979-8-4007-0159-7/23/08}

\begin{document}

\begin{teaserfigure}
\includegraphics[page=1,width=\linewidth,trim={0 4.3cm 0 0}]{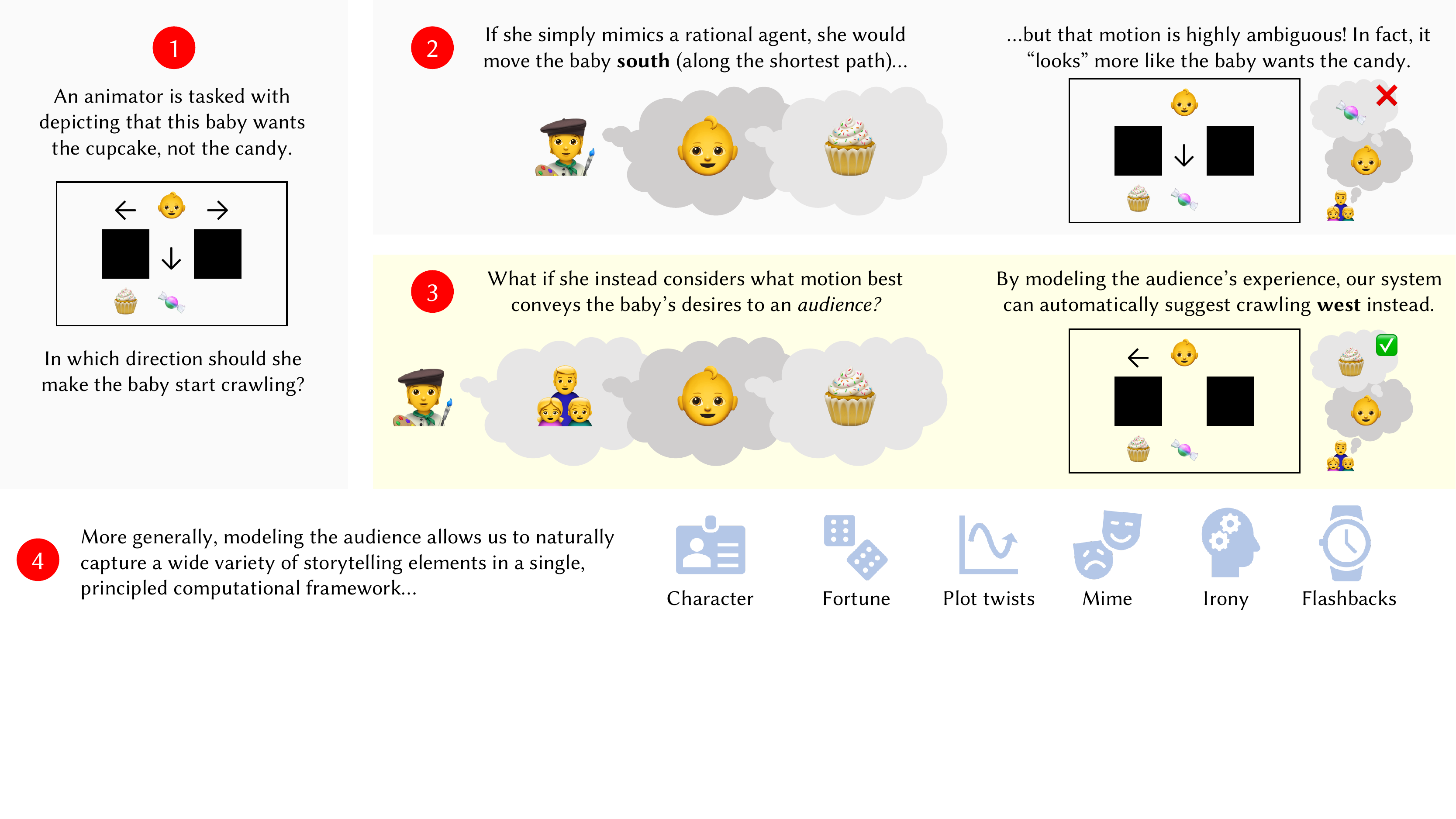}
\end{teaserfigure}

\begin{abstract}
Great storytellers know how to take us on a journey. They direct characters to act---not necessarily in the most rational way---but rather in a way that leads to interesting situations, and ultimately creates an impactful experience for audience members looking on.

If audience experience is what matters most, then can we help artists and animators \emph{directly} craft such experiences, independent of the concrete character actions needed to evoke those experiences? In this paper, we offer a novel computational framework for such tools. Our key idea is to optimize animations with respect to \emph{simulated} audience members' experiences. To simulate the audience, we borrow an established principle from cognitive science: that human social intuition can be modeled as ``inverse planning,'' the task of inferring an agent's (hidden) goals from its (observed) actions. Building on this model, we treat storytelling as ``\emph{inverse} inverse planning,'' the task of choosing actions to manipulate an inverse planner's inferences. Our framework is grounded in literary theory, naturally capturing many storytelling elements from first principles. We give a series of examples to demonstrate this, with supporting evidence from human subject studies.
\end{abstract}

\maketitle


\section{Introduction}
\label{sec:introduction}

The ACM SIGGRAPH vision statement is ``enabling everyone to tell their stories.'' But why do stories need to be ``told''? Why is a well-told story more impactful than the everyday events occurring all around us?
The answer is that story\emph{telling} is done for an audience: unlike the dull, indifferent progression of real life, stories are propelled by deliberate artistic choices made solely to craft the audience's experience. Storytellers can intentionally build suspense by withholding information, or cause a surprise ``twist'' by revealing it at the right time---and they can violate the laws of physics at will.

\begin{table*}
\caption{``Inverse inverse planning'' is analogous to past graphics work on ``inverse inverse rendering.''}\label{tab:i2p}
\begin{tabular}{p{7.0cm}p{5cm}p{5cm}}
\toprule
 & \textbf{Static images} & \textbf{Stories and animations} \\ \midrule
World model (reality $\rightarrow$ stimulus) & Simulate light transport by rendering & Simulate goal-driven agent by planning \\
Audience model (stimulus $\rightarrow$ percept of reality) & Inverse rendering & Inverse planning \cite{baker2009action} \\
Depiction task (desired percept $\rightarrow$ synthetic stimulus) & Inverse inverse rendering & Inverse inverse planning \\
Theoretical proposal & \citet{durand2002perceptual} & \citet{kukkonen2014bayesian} \\
Concrete computational model & \citet{chandra2022designing} & \textbf{This paper} \\
\bottomrule
\end{tabular}
\end{table*}

This insight suggests a novel declarative interface for computational storytelling. Instead of authoring a low-level sequence of events, \textbf{what if a storyteller could directly specify the high-level \emph{desired audience experience?}} The computer would then automatically synthesize a concrete animation that evokes that experience.
In this paper, we present precisely such an algorithm. We optimize animations to create the desired experience in \emph{simulated} audience-members, modeled computationally using principled frameworks grounded in modern cognitive science. Because these computational models are well-tuned to human intuitions, the optimized animations create the desired effect in human audiences.

What models should we use? A long line of work from the computational cognitive science community, originating with \citet{baker2009action}, has modeled human social cognition with \emph{Bayesian inference}. This model posits that we always have uncertainty about the goals of people we see around us---however, their actions reveal information about those goals. For example, if the baby in the figure above moves south, we get the sense it probably wants the candy, though it might also want the cupcake.
%
If we model people as approximately-rational planners, where $\p(\text{action} \mid \text{goal})$ is highest for the best action to achieve a goal, then action understanding is \emph{inverse planning:} the problem of inferring $\p(\text{goal} \mid \text{action})$.

In this paper, we propose that storytellers add another recursive layer: they choose actions to persuade inverse-planners of certain goals.
An actor portraying the baby might choose to move \emph{west} to emphasize that it wants the cupcake, not the candy---even though moving west is a sub-optimal choice for an audience-agnostic rational agent.
More generally, \textbf{just as \citeauthor{baker2009action} cast ``action understanding as inverse planning,'' we cast ``\emph{acting} as \mbox{\emph{inverse}} \mbox{inverse} planning.''} This framework is remarkably flexible: it provides a single unifying language---optimization over Bayesian inference---in which storytellers can express a variety of storytelling tasks from first principles. For example, a storyteller might ask the system for $\argmax_\text{action} \p(\text{baby wants cupcake} \mid \text{action})$. The optimizer could then automatically suggest \emph{west} as the best solution. \textbf{More broadly, our model naturally captures elements of character, setting, plot, irony, flashbacks, and more.}

``Inverse inverse planning'' is closely related to the graphics community's past work in depiction of static scenes. In a seminal SIGGRAPH course, \citet{durand2002perceptual} abstractly cast the task of visual ``depiction'' as ``inverse inverse rendering.'' An artist paints the canvas to cause viewers (``inverse renderers'') to infer the scene they seek to convey, even if the resulting painting is not a faithful, physically-accurate rendering of that scene. For example, a caricature exaggerates facial features to emphasize the subject's identity to a viewer. Recently, \citet{chandra2022designing} concretely applied ``inverse inverse rendering'' to produce new perceptual experiences (illusions) by optimizing over Bayesian inverse-rendering models of human vision. Here, we extend the same framework to animation, which is often called the ``illusion of life'' \cite{thomas1995illusion}. We illustrate the analogy between these lines of work in Table~\ref{tab:i2p}.

Inverse inverse planning is also grounded in ideas from literary theory. For example, \citet{kukkonen2014bayesian} abstractly casts storytelling as ``probability design,'' thinking of narratives as sequences of observations presented to a Bayesian audience. We provide the first concrete computational evidence supporting that vision.
Similarly, American writer Kurt Vonnegut's master's thesis (unpublished, but see the \href{https://www.youtube.com/watch?v=oP3c1h8v2ZQ}{linked lecture video}) argues that all stories have simple ``shapes'' defined by the trajectory of the protagonist's fortunes over time \cite{vonnegut2011man, kiley2016game}. \citet{reagan2016emotional} apply statistical analyses to a corpus of novels to extract representative story shapes, and suggest that future work investigate the ``opposite direction'' of generating a story that matches a given dramatic arc. Our work here takes a first step in that direction (Section~\ref{sec:app-shapes}).

\textbf{Our core contribution in this paper is a novel cognitively-inspired formalism for storytelling and animation in computer graphics.}
We begin by reviewing how cognitive scientists treat ``action understanding as inverse planning,'' outlining a concrete model proposed by \citet{ullman2009help} (Section~\ref{sec:review}). Next, we show how to implement ``\emph{inverse} inverse planning'' on top of this model, and demonstrate the remarkable flexibility of our system with a wide variety of example applications (Section~\ref{sec:applications}). We then offer some evidence from IRB-approved human subject studies, showing that inverse inverse planning is more effective than audience-agnostic ``na\"ive'' planning for depiction tasks (Section~\ref{sec:experiments}). For example, in one study viewers were $9\times$ likelier to correctly discern the relationship between two characters when presented with animations created by our method, as compared to na\"ive planning baselines. Finally, we discuss additional related work (Section~\ref{sec:related-work}) and conclude with prospects for future work (Section~\ref{sec:future-work}). Sample code for implementing our examples is available in the supplementary materials and online at \mbox{\url{https://people.csail.mit.edu/kach/a2i2p/}}.

\section{Inverse planning}
\label{sec:review}

Humans have an astonishing intuitive ability to make inferences about other intelligent agents. Consider the well-known 90-second animation by psychologists \citet{heider1944experimental}, which shows three simple shapes moving in 2D. Although these simple geometric shapes do not look like humans, nearly everyone attributes complex human traits to them: viewers consistently report that the shapes have certain beliefs, desires, emotional states, and moralities. A long line of work from the cognitive science community, originating with \citet{baker2009action, baker2007goal}, has modeled this intuition with Bayesian inference. These models posit that when we observe an agent take an action, we infer that agent's goal by applying Bayes' rule: $\p(\text{goal} \mid \text{action}) \propto \p(\text{action} \mid \text{goal})\p(\text{goal})$. Here, the likelihood $\p(\text{action} \mid \text{goal})$ can be estimated by imagining the agent's planning process: the more optimal an action is for the agent, the likelier it should be. $\p(\text{goal})$ reflects our prior beliefs about goals.

This model is considered ``inverse planning'' in the sense that while \emph{planning} outputs a plan for a goal, \emph{inverse planning} infers a goal from a plan. Inverse planning has been extended in a variety of directions, such as in modeling agents that reason about each other via a ``theory of mind'' \cite{ullman2009help, baker2008theory, tauber2011using}, and agents who engage in human-like planning, which is not always optimal \cite{zhi2020online}. Recent work has shown that inverse planning can account for human judgements of human kinematic motions \cite{qian2021modeling}, judgements made by young children \cite{pesowski2020children}, and actions humans take to influence each other \cite{ho2021communication, ho2022planning, shafto2014rational, radkani2022modeling, goodman2016pragmatic, yoon2016talking}.

For much of this paper, we will discuss a concrete world model proposed by \citet{ullman2009help}, which has been tuned to match human intuitions and demonstrated to do so via human subject studies. In the rest of this section, we briefly review this world model and how inverse planning functions in it.

\begin{figure*}[h]
\includegraphics[page=12,width=\linewidth,trim={0 4.63in 0 0},clip]{Figures.pdf}
\includegraphics[width=0.9\linewidth,trim={0 0.3in 0 0}]{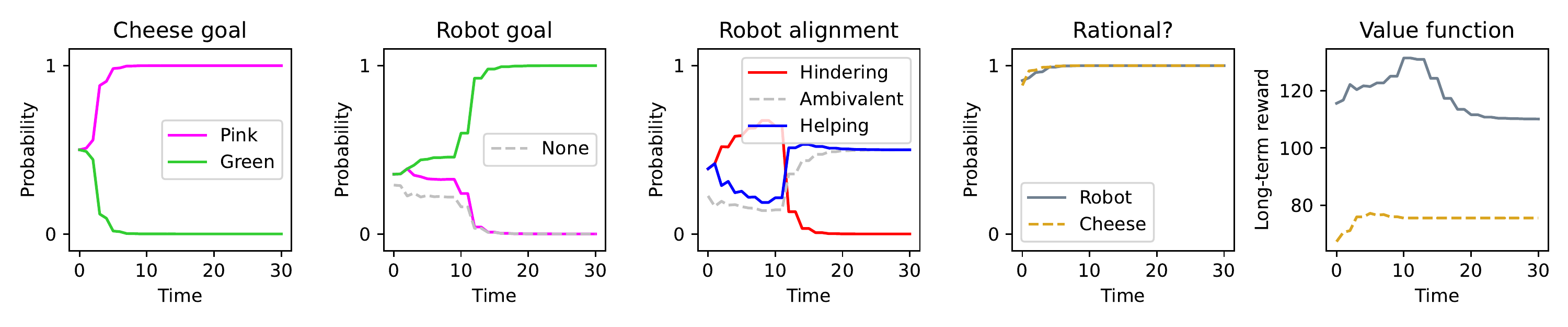}
\caption{\textbf{(top)} Suppose we animate a robot that is helping the cheese reach its goal, by having both characters follow their optimal policies (``na\"ive planning,'' Section~\ref{sec:review}). \textbf{This produces a poor depiction:} it is not clear that the robot wants to \emph{help} the cheese, only that it wants to go to green. \textbf{(bottom)} The inverse planner agrees. It infers that the cheese wants pink (first plot), and that the robot wants green (second plot; notice bump at $t=10$ when the robot reaches green and stays). But it remains uncertain about the robot's alignment (third plot), because the robot's behavior is consistent with both ambivalence to the cheese and wanting to help the cheese (but doing nothing because the cheese is already at pink). Can we do better? Yes: by \emph{inverse inverse planning!} (Figure~\ref{fig:grid-i2p}).}\label{fig:grid-plan}
\end{figure*}

\begin{figure*}[h]
\includegraphics[page=2,width=\linewidth,trim={0 4.8in 0 0},clip]{Figures.pdf}
\includegraphics[width=0.9\linewidth,trim={0 0.3in 0 0}]{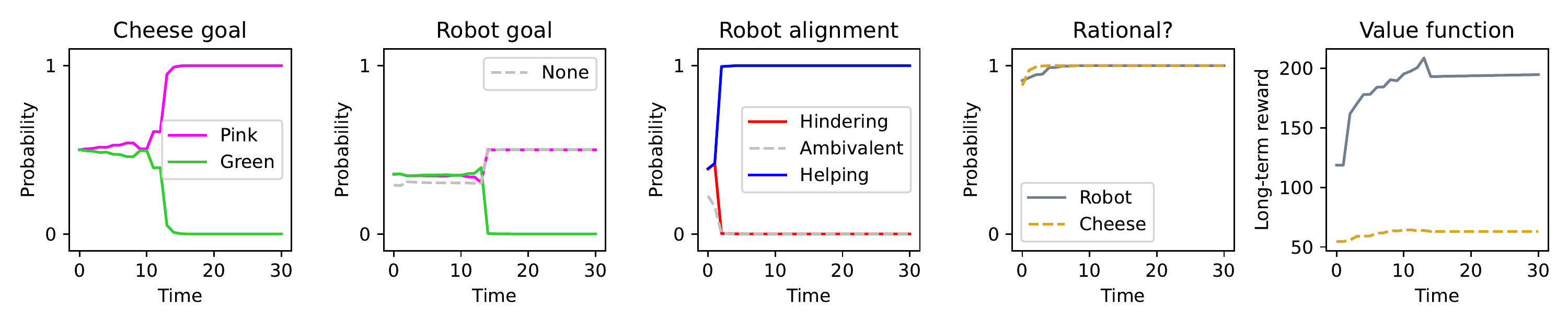}
\caption{\textbf{(top)} Using inverse inverse planning, we optimize animations that maximize the inverse planner's belief that the robot is helpful (Section~\ref{sec:applications}). This finds a much more effective depiction. \textbf{(bottom)} Now, it is clear that the robot is helping, and indeed the model is confident of that from the beginning (third plot).}\label{fig:grid-i2p}
\end{figure*}

\paragraph{World model}
\citeauthor{ullman2009help}'s world (Figure~\ref{fig:grid-plan}) consists of two agents moving in a maze on a grid. In our animations, we will stylize them as two characters: a \emph{robot} and an \emph{enchanted animate cheese cube} in a kitchen.
The two agents can move north / south / east / west through the kitchen or stay in place. The cheese is ``weak'' and only succeeds in moving 60\% of the time. However, the robot is ``strong'' and can push the cheese cube along. A \emph{table} in kitchen blocks the cheese's motions, but can be moved by the robot. Finally, the kitchen floor has two special tiles, \emph{pink} and \emph{green}.

The two characters can have a variety of natural goals in the kitchen. The cheese and the robot could each ``want'' to sit on either the pink or green tile (think of the tiles as ``MacGuffins''). Additionally, because the robot is strong enough to move the cheese by pushing it around, it could want to ``help'' or ``hinder'' the cheese from reaching its goal.

\paragraph{Planning} These dynamics and goals can be formalized as a multi-agent Markov Decision Process or MDP (for a detailed introduction to MDPs, we refer readers to a recent textbook by \citet{kochenderfer2022algorithms}). The state space $\mathcal{S}$ encodes the positions of the robot, cheese, and table in the kitchen. The action space $\mathcal{A}$ for each agent is $\{\leftarrow, \rightarrow, \uparrow, \downarrow, \text{stay}\}$. The transition function for each agent encodes how each action affects the state as described above (the transition function for the cheese is stochastic because the action may fail).

Finally, the reward function for each agent captures the agents' goals. The cheese and the robot each receive a fixed reward if they are on their respective goal tiles (pink or green), and pay a small cost for moving instead of staying in place. In addition, the robot receives a ``social reward'' based on the cheese's reward on this turn. Specifically, if the cheese earns reward $r_\text{cheese}$, then the robot earns a bonus reward $\rho_\text{robot}\cdot r_\text{cheese}$ where $\rho_\text{robot}$ is positive if the robot is helping, negative if the robot is hindering, and zero if the robot is neutral. $\rho_\text{robot}$ can be chosen from $\{-3, -1, 0, +1, +3\}$, expressing the range from ``highly adversarial'' to ``indifferent'' to ``highly helpful.''

Having written this concrete reward function, \citeauthor{ullman2009help} compute optimal policies for the two agents by running value iteration \cite{bellman1966dynamic}, setting $\gamma=0.99$ and extracting $Q$-functions from the computed value functions. They use a hierarchical softmax strategy, first computing a policy for the cheese assuming the robot moves uniformly at random, and then computing a policy for the robot assuming the cheese selects actions with probabilities given by the softmax of its $Q$-function with temperature $\beta=2.0$. This allows for two recursive levels of ``theory of mind'' in the planner: the robot models the cheese modeling the robot.

Figure~\ref{fig:grid-plan} shows a sample animation we generated by running these optimal policies na\"ively for a helpful robot in a random scenario (i.e. in a random state, with random goals where $\rho_\text{robot} > 0$). As we argue in the caption, this is a poor depiction of helpfulness---instead, it inadvertently conveys indifference. Next, we will quantitatively capture why the depiction fails in this way, by using \emph{inverse planning} to model how humans experience these animations.

\paragraph{Inverse planning} \citeauthor{ullman2009help}'s inverse planner makes inferences about agents' (hidden) goals from (observed) actions. Let a \emph{hypothesis} be a tuple: \[
H=\left\langle
G_\text{cheese} \in \{p, g\},
G_\text{robot} \in \{p, g, \emptyset\},
\rho_\text{robot} \in \{0, \pm1, \pm3\}
\right\rangle
\] Above, we showed that for fixed $H$, we can use value iteration to compute $Q^H_\text{robot}(s, a)$ and $Q^H_\text{cheese}(s, a)$ for state $s \in \mathcal{S}$ and action $a \in \mathcal{A}$. Assuming the softmax-rational model above, this induces a probability distribution over each character $c$'s actions:
\begin{equation}\label{eqn:likelihood}
\p(s \rightarrow a \mid H) \propto \text{exp}\left(\beta \cdot Q^H_c(s, a)\right)
\end{equation}
Now, if we observe an agent take action $a$ from state $s$, we can apply Bayes' rule to calculate the probability of the characters' goals:
\begin{equation}\label{eqn:bayesrule}
\p(H \mid s \rightarrow a) \propto \p(s \rightarrow a \mid H)\p(H)
\end{equation}
Here, $\p(s \rightarrow a \mid H)$ is the \emph{likelihood} given by Equation~(\ref{eqn:likelihood}) and $\p(H)$ is our prior belief about the characters' goals. Before the animation plays, we assume a uniform prior over $H$, i.e. $\p(H) \propto 1$, reflecting our ignorance about the characters' goals. Each time we observe a character take an action, we \emph{update} our belief about hypothesis $H$.
To implement this inference, we maintain a mapping from hypotheses to probabilities. When we observe an action, we update the conditional probability of each hypothesis using Equation~(\ref{eqn:bayesrule}), re-normalizing so the distribution sums to 1.

We can see inverse planning in action in the bottom of Figure~\ref{fig:grid-plan}. The first three plots show how $\p(G_\text{cheese})$, $\p(G_\text{robot})$, and $\p(\rho_\text{robot})$ change as the animation plays and more actions are observed (plots \#4 and \#5 are introduced in Section~\ref{sec:details}). As expected, the model is not confident that the robot is helping. In the next section, we show how to use \emph{inverse inverse planning} to create animations that \emph{do} effectively depict helping (and other scenarios).

\section{\emph{Inverse} inverse planning}
\label{sec:applications}

Now that we can model an audience's experience of an animation by Bayesian \emph{inverse planning}, we can create new animations by \emph{inverse inverse planning}---that is, by optimizing over Bayesian inference.

To be precise, we optimize over \emph{scripts} of length $T$, where a script $\sigma$ is given by an initial state $s_0$ and a sequence of valid \emph{transitions} $\sigma_t = \langle a_t^\text{robot}, a_t^\text{cheese}, s_t \rangle$ for $1 \leq t \leq T$. Note that $s_{t}$ is not uniquely determined by $s_{t-1}$ and $\langle a_t^\text{robot}, a_t^\text{cheese} \rangle$ because state transitions may be non-deterministic. For example, recall how the cheese only succeeds in moving with probability 0.6---we allow our optimizer to choose whether or not the move succeeds.

Suppose, like in the previous section, that we wanted an animation that depicts the robot as helping the cheese. We can express this task in a simple and natural objective function over scripts:
\[
f_\text{help}(\sigma) = \sum_{1 \leq t \leq T} \p(\rho_\text{robot} > 0 \mid \sigma_{1:t})
\]
The objective $f_\text{help}$ is maximized for scripts where at every time $t$, based on observing the animation up to time $t$ (i.e. $\sigma_{1:t}$), a viewer has a strong belief that the robot is helping (i.e. $\rho_\text{robot} > 0$).

Notice that $f_\text{help}$ does not say anything about $G_\text{cheese}$, $G_\text{robot}$, or even the initial positions of the characters in $s_0$. If we were instead making animations by simulating optimal agents, we would have to specify all of these parameters up-front, even though they are unrelated to the abstract goal of depicting ``helping.'' In this way, inverse inverse planning allows for modular, high-level reasoning about the essence of a story in a way that planning itself does not.

To optimize scripts for a given objective like $f_\text{help}$, we use beam search. We sample a set of random initial states to seed our candidate scripts. For each candidate script, we independently run beam search over transition sequences, where the search heuristic is the objective applied to the current script ``prefix.'' The number of initial states sampled and the beam width are hyperparameters of the algorithm. Unless otherwise noted, for the examples below we used 500 states with beam width 1 (i.e. greedy search).

Aside from some smaller details (see Section~\ref{sec:details}), this is all we need to inverse inverse plan. When we run the optimizer on $f_\text{help}$ with $T=15$, we get a rendered animation within just a couple of minutes (Figure~\ref{fig:grid-i2p}). The cheese moves towards pink, and the robot pushes it along. Upon watching this animation, a rational viewer would infer that the cheese wanted pink (because of its initial motion towards pink), and that the robot wanted to help the cheese (because it pushed the cheese to pink and stepped back afterwards).
This animation is a significantly more effective depiction of helping than the one we generated earlier by na\"ive planning (Figure~\ref{fig:grid-plan}).

\subsection{Applications}

We now show a wide variety of additional examples of encoding classic storytelling elements as inverse inverse planning. We encourage readers to imagine how they would depict these scenes themselves before looking at our system's outputs, available in the supplementary video (timestamps in text) and summarized in Figures~\ref{fig:example-animations-i} and \ref{fig:example-animations-ii}. All outputs shown are from the same random seed (0). Simple variations emerge with different seeds. Optimization and rendering takes less than two minutes on a server with 44 CPUs.

\subsubsection{Character}
\label{sec:app-character}

Above, we showed how to depict the character of a helpful robot. Similarly, we can ask for an animation of a \emph{hindering} robot.
In the generated animation, the cheese first moves to green. Then the robot pushes it into a corner and blocks the way to green (Fig~\ref{fig:example-animations-i}A/1:40). In comparison, regular planning from a random initial state yields an animation where the robot moves to green, blocking the cheese. To an observer it is unclear whether the robot is intentionally hindering, or indifferent to the cheese and wanting green for itself (Fig~\ref{fig:example-animations-i}B/1:16). Inverse inverse planning avoids this ambiguity because the robot never moves onto green.

\subsubsection{Plot twists}

We next consider the ``plot twist,'' a storytelling device where an unexpected event radically alters the audience's expectations. For example, a classic plot twist reveals that a seemingly friendly character was adversarial all along. Here, we ask for an animation where the robot \emph{appears} to be helpful at first, but at $t=T/2$ is revealed to be hindering instead.
\begin{align*}
f_\text{twist}(\sigma) = \sum_t & \begin{cases}
\p(\rho_\text{robot} > 0 \mid \sigma_{1:t}) & t \leq T/2 \\
\p(\rho_\text{robot} < 0 \mid \sigma_{1:t}) & t > T/2
\end{cases}
\end{align*}
In the generated animation, the robot ``helpfully'' pushes the cheese to pink. However, upon reaching pink it continues pushing, trapping the cheese along the wall. This surprising action reveals that the robot's true intention was to hinder all along (Fig~\ref{fig:example-animations-i}C/2:39).
We can also ask for the reverse, a video where the robot appears to be hindering but was helping all along.
%
%
Now, the cheese moves to pink and the robot approaches as if to push it off (hindering). However, the cheese continues moving, revealing that it wanted green all along. The robot helpfully pushes it there (Fig~\ref{fig:example-animations-i}D/3:14).

\subsubsection{Irony}

Next, we consider \emph{dramatic irony}, which occurs when the audience has a different understanding of a situation than the characters in that situation. Because our system explicitly models the audience's understanding, we can straightforwardly express scenes with dramatic irony. Here, we design an objective function for scenes where the robot appears to be trying to help, but mistakenly hinders because of its false belief about the cheese's goal. We use conditional probability to express that the robot should appear to be helpful \emph{if} the cheese had a different goal.
\begin{align*}
f_\text{irony}(\sigma) = \sum_t & +\p(G_\text{cheese}=\text{green} \mid \sigma_{1:t}) \\
&+ \p(\rho_\text{robot} < 0 \mid \sigma_{1:t}, G_\text{cheese}=\text{green}) \\
&+ \p(\rho_\text{robot} > 0 \mid \sigma_{1:t}, G_\text{cheese}=\text{pink})
\end{align*}
In the generated animation, the cheese moves to green, but the robot pushes it off and towards pink. When the cheese tries to move back, the robot ``helpfully'' guides it back to pink (Fig~\ref{fig:example-animations-i}E/3:36).

\subsubsection{Flashbacks} Nonlinear discourse is a storytelling technique where information in a story is revealed out of chronological order. For example, a ``flashback'' can be used to re-contextualize a scene, giving it heightened significance or new meaning. Here, we show an example of using inverse inverse planning to design flashbacks.

Imagine we saw a glimpse of the robot pushing the cheese east, away from the pink and green goals. Can we show a flashback animation that casts this action as helping? Let $c(\sigma)$ be the script $\sigma$ with a single transition appended, in which the robot moves east while the cheese stays. We can now apply the objective function for ``helping'' over $c(\sigma)$, with an additional cost function that ensures that the robot pushes the cheese when it moves east (i.e. the cheese is directly to the east of the robot).
\[
f_\text{flashback-help}(\sigma) = f_\text{help}(c(\sigma)) + \begin{cases}
1 & \text{cheese pushed at }t=T+1 \\
0 & \text{else}
\end{cases}
\]
For this example, we raise the beam width to 100 because we observed that greedy search was easily trapped in local minima. In the generated flashback, the cheese is trying to go all the way around the room to pink because the table is blocking a door along the shorter path. This casts the robot's pushing as helping (Fig~\ref{fig:example-animations-ii}G/4:03).
Of course, we can instead substitute $f_\text{hinder}$ to find a flashback that casts the action as hindering.
In the generated flashback, the cheese tries to move directly to pink, casting the robot's push as hindering (Fig~\ref{fig:example-animations-ii}H/4:18).

\subsubsection{Narrative arc}\label{sec:app-shapes}

Recall from the introduction Vonnegut's theory that the ``shape'' of a story is the trajectory of the main character's fortunes. We would like to optimize for animations where the robot's fortunes decline and then rise again, creating a ``story arc.''

To heighten the effect, we add a mechanism for characters' fortunes to change based on external events. Since ancient times, storytellers have propelled or resolved plots by introducing a new element from outside the world of the story. For example, a heavenly chariot descends from the sky to save the characters in Euripides' tragedy \emph{Medea}. Literary theorists refer to this pattern as ``deus ex machina,'' because gods and divine interventions (``deus'') were once lowered onto stages with mechanical contraptions (``ex machina''). We create the possibility for ``deus ex machina'' by creating a special type of transition $\langle \text{deus}, x, y \rangle$ where the obstructing table ``falls from the sky'' into the kitchen at position $(x, y)$. This transition can only be used once, and only in worlds where the table is not already present. Note that the characters' learned policies do not account for the possibility of this transition occurring; nor do audiences know to expect it---it is a surprise from ``outside'' the fictional world.

With this enhancement, we can search for stories where the value function of the robot (learned by value iteration) correlates with the rise-fall-rise of 1.5 periods of a sinusoid:
\begin{align*}
f_\text{arc}(\sigma) = \sum_t  & + \sin(t/T\cdot 3\pi) \cdot \E\left[V^H_\text{robot}(s_t) \mid \sigma_{1:t}\right] \\
& -0.1\cdot D_\text{KL}(H_{1:t-1} \parallel H_{1:t})
\end{align*}
We do not specify anything else about the story. However, we introduce a new term to enforce that the characters' goals are consistent. Otherwise, we might get stories where the robot's apparent fortune changes because its apparent goal changes. To enforce this consistency, we minimize the KL-divergence between our beliefs about the characters' goals before and after each observed action. Here, $H_{1:t}$ is a random variable with probability distribution $\p(H \mid \sigma_{1:t})$.

In the generated animation, the robot starts helping the cheese to pink. However, the table falls onto pink at the last moment. Then the robot moves the table out of the way, allowing the cheese to finally reach pink (Fig~\ref{fig:example-animations-i}F/4:32).

\subsection{Implementation details}
\label{sec:details}

To get high-quality results, we need to account for some storytelling-specific details that \citeauthor{ullman2009help} did not need to address.

First, we would like characters to appear rational.
\citeauthor{ullman2009help}'s model presupposed that agents in the animations were acting rationally, because it was only tested on hand-designed stories featuring rational agents. However, we run our model on arbitrary scripts with potentially-irrational behavior. Thus, we need to add hypotheses for irrational agents. We augment our hypothesis tuple with two boolean variables $R_\text{robot}$ and $R_\text{cheese}$, which track whether each character is rational, positing in the likelihood function that irrational agents select actions uniformly at random. Additionally, when the cheese is irrational, we only include hypotheses where the robot is indifferent to it.
Now that we can infer the rationality of each character, we automatically add a rationality term,
\(
f_\text{rational}(\sigma) = \sum_t \p(R_\text{robot} \wedge R_\text{cheese} \mid \sigma_{1:t})
\), to the artist-provided objective function. Additionally, we implicitly condition the artist's probability calculations on $R_\text{robot} \wedge R_\text{cheese}$ because they likely have rational characters in mind when designing the story.

Similarly, we must ensure that the environment behaves plausibly (\citeauthor{kukkonen2014bayesian} calls this \emph{versimilitude}). Recall that the cheese only succeeds in moving with probability 0.6. Our videos must faithfully reflect this. If we show the cheese try and fail to move ten times consecutively, audiences would either dismiss the video as implausible and contrived, or update their belief about the success probability. We avoid such pathological cases by introducing an environmental consistency term
\(
f_\text{env}(\sigma) = -(\hat{p}(\sigma) - 0.6)^2
\)
where $\hat{p}(\sigma)$ is the proportion of times the cheese successfully moves in $\sigma$.

Finally, we must address the possibility that goals change over time. An observer might also discount or simply forget past evidence. For example, if the robot acted adversarially in the first part of an animation, but then sat still for a long time, we might be uncertain whether it is still adversarial. If it then begins moving irrationally, we would immediately perceive it as irrational, discounting past rational behavior. Following \citet{baker2009action}, we account for this by positing that after each turn, with some small probability $\epsilon=10^{-5}$, all latent variables are reassigned uniformly at random. This is easily implemented by adjusting $\p(H) \mapsto \p(H)\cdot(1-\epsilon) + (1- \p(H))/(N-1)\cdot\epsilon$, where $N$ is the number of hypotheses. This adjustment discounts past evidence, forcing the inverse inverse planner to continue providing new evidence throughout the story.

\subsection{Miming in a physics world}\label{sec:app-mime}

Finally, we depart from this grid-world setting and move to a more naturalistic physics-based setting. We were inspired by the short animated film \emph{Sisyphus} \cite{jankovics1974sisyphus}, which depicts the character Sisyphus from Greek mythology pushing a heavy boulder up a hill. By exaggerating Sisyphus' movements, the animation creates a dramatic impression of the boulder's immense weight. We wondered if inverse inverse planning could evoke such an effect: that is, make a character ``mime'' a heavy object.

To model the scenario, we created a small physics-based environment consisting of a mass-spring system, and used it to design a ``Luxo lamp''-style hopper similar to that of \citet{witkin1988spacetime}. We attached the hopper to a box on a hill (Figure~\ref{fig:lampworld}).

For \emph{planning}, we built a differentiable physics simulator for this environment and used it to train a controller to pull the box up the hill using the Short-Horizon Actor-Critic algorithm \cite{xu2022accelerated}. Actor-critic algorithms jointly train two neural networks: a policy $\pi(s; \theta)$ that computes actuations for the agent at state $s$, and a value function $V(s; \phi)$ that computes the optimal-long term reward attainable from $s$. The learned parameters of these neural networks are $\theta$ and $\phi$. We optimized actor-critic pairs for two box weights, light (0.1) and heavy (0.5) to obtain $\theta_{\{0.1, 0.5\}}$ and $\phi_{\{0.1, 0.5\}}$.

Next, we used \emph{inverse planning} to model a viewer's impression of the box weight. Following \citet{battaglia2013simulation}'s work on Bayesian models of human physics perception, we compared each frame of the observed trajectory against hypothetical simulations of the physical system using $\theta_{0.1}$ and $\theta_{0.5}$.

Finally, we created videos of the hopper ``miming'' pulling a heavy box with \emph{inverse inverse planning}. We used gradient descent to optimize a trajectory that maximizes the inverse planner's confidence that the box is heavy (even though the box was actually light in the simulator). We parameterized trajectories by residuals over the optimal policy $\theta_{0.1}$: for each time $t$ we optimize a residual actuation that is added to the optimal actuation given by $\pi(s_t; \theta_{0.1})$. This results in the hopper ``pretending'' to struggle as it pulls the box (supplementary video, 5:30). Our human subject studies (Section~\ref{sec:lamp-exp}) confirm that the hopper indeed convinces viewers that the box is heavier than it truly is.

\subsubsection{Stumbling}\label{sec:app-stumble}

We additionally replicate the ``shapes of stories'' example from Section~\ref{sec:app-shapes} in this domain. We use gradient descent to optimize a trajectory in which value function $V(s_t; \phi_{0.5})$ dips from time $t_s$ to time $t_f$:
\[
f_\text{arc}(s) = \sum_t V(s_t; \phi_{0.5}) \cdot \begin{cases}
-1 & t_s \leq t \leq t_f \\
+1 & \text{else}
\end{cases}
\] As before, we optimize residuals over the optimal policy $\theta_{0.5}$. The resulting animation shows the hopper ``stumble'' at time $t_s$ and ``recover'' at time $t_f$ (Figure~\ref{fig:lamp-stumble}; supplementary video, 7:05).

\subsection{Human subject studies}
\label{sec:experiments}

We empirically evaluated our method on both the ``kitchen'' and ``hill'' domains. Our guiding question was: \emph{Is our inverse inverse planner more effective at depicting desired conditions than a regular na\"ive planner?} To investigate this question, we designed two experiments.

\subsubsection{Kitchen}\label{sec:grid-exp}
We sought to answer whether inverse inverse planning better depicts the robot's relationship with the cheese than na\"ive planning. We generated 20 animations each of helping, hindering, and indifference (i.e. $\rho_\text{robot}=0$) using random seeds 0-19, using both inverse inverse planning and na\"ive planning. We recruited 98 online participants (English-speaking, 80\% male, average age 40, min 18, max 71) and showed each participant a random shuffled subset of these animations. Note that participants were not aware that the videos could have come from two different algorithms, or even that they were computer-generated at all. For each video, we asked participants to report whether the robot was helping the cheese, hindering it, indifferent to it, or whether the animation was unclear. We measured the proportion of responses that matched the desired depiction target.

\begin{figure*}
\centering
\includegraphics[width=\linewidth]{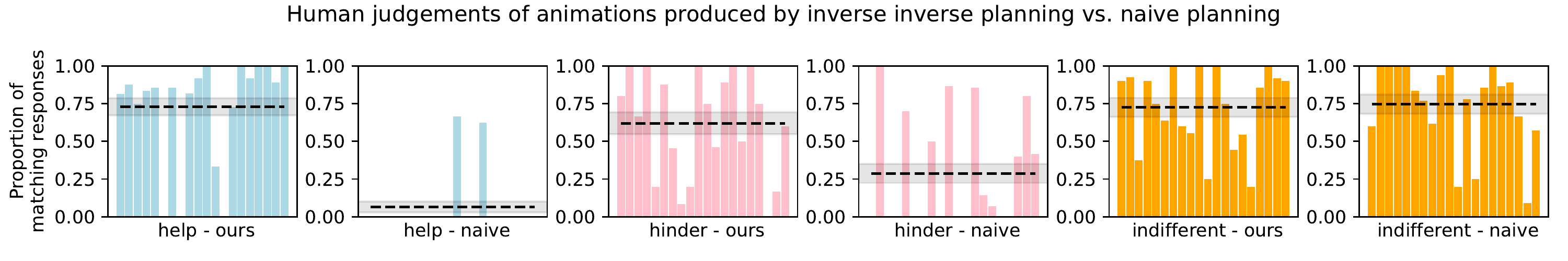}%
\caption{In Section~\ref{sec:grid-exp} we empirically compare inverse inverse planning and na\"ive planning on three depiction tasks: showing the robot to be helping, hindering, or indifferent. Participants viewed the resulting animations and reported their impression of the robot (help / hinder / indifferent / unclear). Each bar represents a separate animation (from a different random seed), showing the proportion of participants who reported the desired response for that animation. Horizontal dashed lines are averages across animations for each condition (higher is better) and shaded areas span 95\% confidence intervals for each condition. \textbf{Our animations are significantly more effective than the na\"ive planning baselines when depicting helping and hindering} ($p \ll 0.01$ for both), and equally effective when showing indifference.}\label{fig:grid-exp}
\end{figure*}

Figure~\ref{fig:grid-exp} shows the results of this experiment. When depicting helping, the average inverse inverse planning animation caused 73\% of viewers to report ``helping,'' while the average na\"ive planning animation only caused 6\% to ($p \ll 0.01$).
When depicting hindering, inverse inverse planning was also significantly better than na\"ive planning (62\% vs. 29\%; $p \ll 0.01$).
Both methods were equivalently effective at depicting indifference (73\% vs. 75\%, n.s.). This is because na\"ive planning animations often have no interaction between the characters, so viewers easily infer indifference.
In summary, we found that inverse inverse planning is indeed better at depicting the robot's relationship with the cheese, especially for conditions that are challenging for na\"ive planning to depict.

\subsubsection{Hill}\label{sec:lamp-exp}
We sought to answer whether the hopper from Section~\ref{sec:app-mime} convincingly ``mimes'' a heavy box. We recruited 35 online participants (English-speaking, 57\% male, average age 34, min 19, max 74) and showed them each a series of pairs of animations. Each animation was randomly either an ``honest'' hopper with a heavy or light box, or a ``mime'' with a light box pretending that the box is heavy. Each video had a different-colored box to emphasize that they may have different weights. Note that participants were not aware that the videos could have come from two different algorithms, or that the hoppers could mime. Participants were asked to select which of the two animations had a heavier box.

\begin{figure*}
\begin{subfigure}[t]{0.20\textwidth}
\centering
\includegraphics[width=\linewidth,trim={3in 1.0in 2.5in 0.5in},clip]{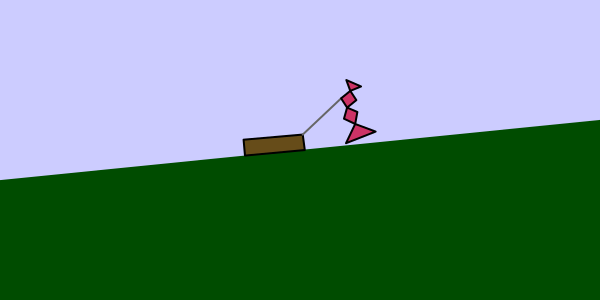}%
\caption{We designed a simple hopper that pulls a box up a hill, and used inverse inverse planning to make it ``mime'' heavy boxes when actually pulling light boxes.}
\label{fig:lampworld}
\end{subfigure}\hspace{0.25cm}
\begin{subfigure}[t]{0.37\textwidth}
\centering
\includegraphics[width=\linewidth]{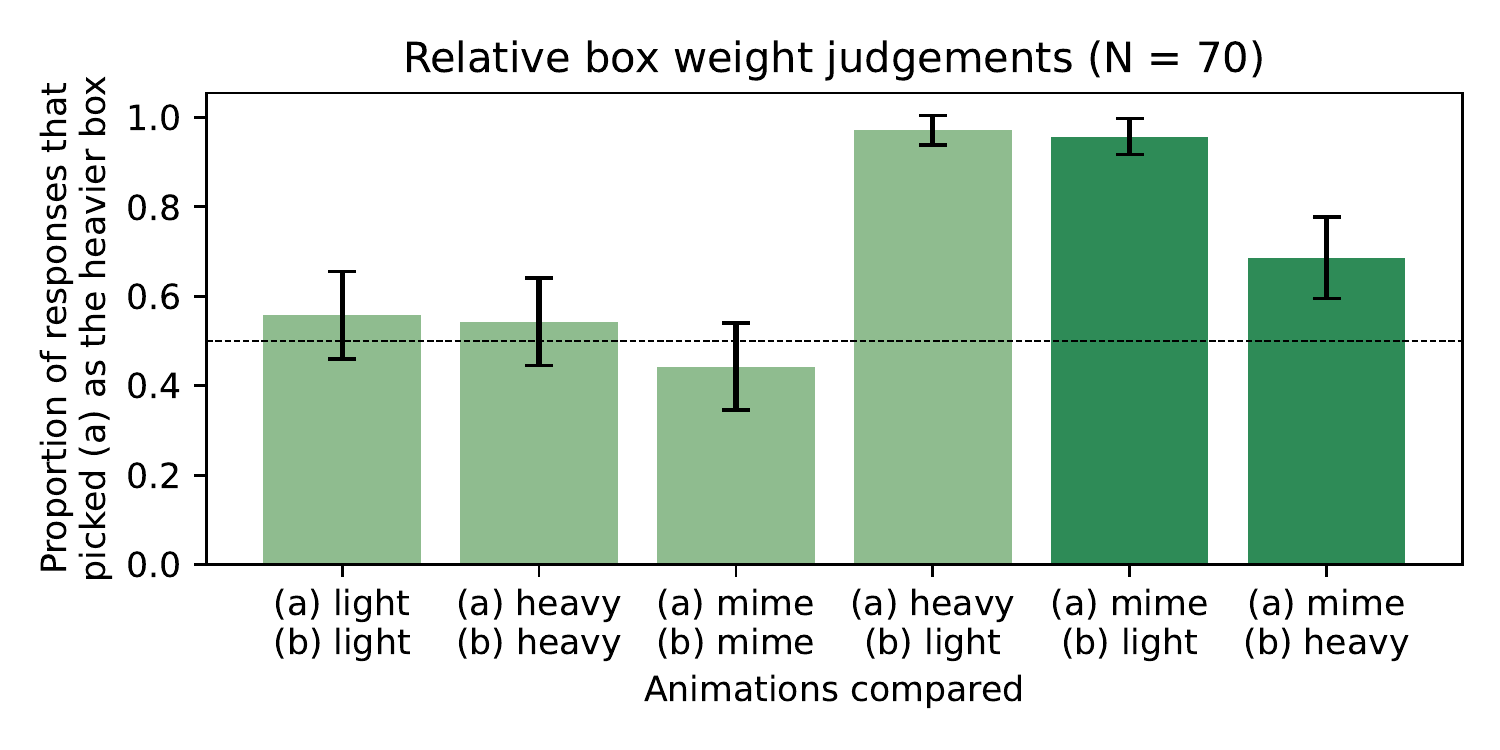}%
\caption{When shown pairs of animations, viewers perceive the ``miming'' hopper from Section~\ref{sec:app-mime} as pulling the heavier box, even though it is pulling a light box (bar \#5, $p \ll 0.01$). This is true even when the \emph{other} box is \emph{actually} heavier (bar~\#6, $p < 0.01$). Error bars show 95\% confidence intervals. (Details in Section~\ref{sec:lamp-exp}.)}
\label{fig:lamp-exp}
\end{subfigure}\hspace{0.25cm}
\begin{subfigure}[t]{0.37\textwidth}
\centering
\includegraphics[width=0.9\linewidth]{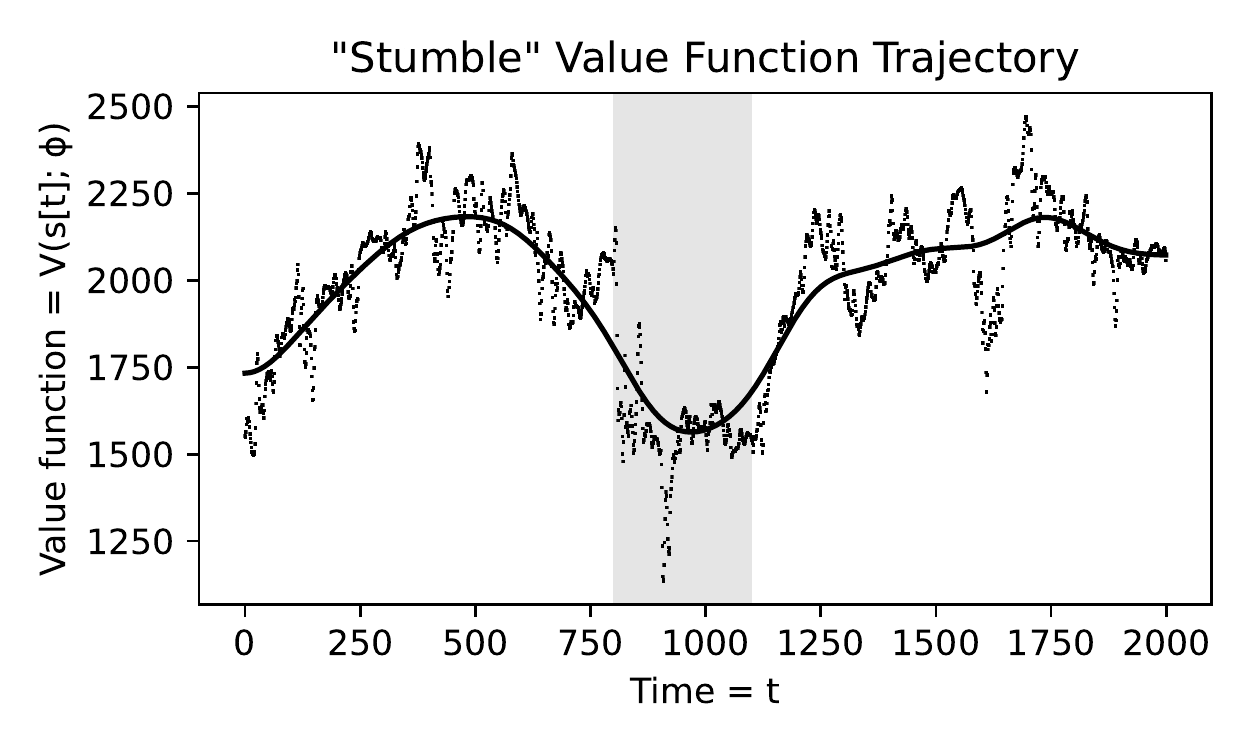}%
\caption{We can force the hopper's value function to ``dip'' from $t_s$ to $t_f$ (the shaded period), causing it to ``stumble'' and then recover on cue (Section~\ref{sec:app-stumble}).}
\label{fig:lamp-stumble}
\end{subfigure}
\caption{Our physics-based world, described in Section~\ref{sec:app-mime}.}
\label{fig:lampworldstuff}
\end{figure*}

Figure~\ref{fig:lamp-exp} shows the results of this experiment. As expected, participants were at chance (50\%) when the animations had the same condition, and between heavy and light boxes they selected the heavy box 97\% of the time ($p \ll 0.01$). The mime convinced 95.7\% of viewers that its box was heavier than the light box, despite being of the same (light) weight ($p \ll 0.01$). Furthermore, it convinced 68.6\% of viewers that it was heavier than the \emph{heavy} box despite being $5\times$ lighter ($p<0.01$). We conclude that the mime successfully convinces viewers that the box is heavier than it truly is.

\section{Additional related work}
\label{sec:related-work}

Section~\ref{sec:review} reviewed related work from cognitive science, on which this paper directly builds. Here we study links to other work in reinforcement learning, computer graphics, and textual storytelling.

\paragraph{Inverse reinforcement learning} The reinforcement learning community has long sought to automatically learn reward functions by ``inverse reinforcement learning'' \cite{arora2021survey, ng2000algorithms, ramachandran2007bayesian}. These methods can be applied to influence observers, like how here we influence an audience's experience of a story. For example, \citet{dragan2015legible} seeks to make a robot's motion ``legible'' to humans by optimizing for an ideal Bayesian observer's ability to predict the robot's intent. Oppositely, \citet{pattanayak2022inverse} use ``inverse-inverse reinforcement learning'' to have agents strategically fool adversarial viewers about their true intentions.

\paragraph{Computer animation} A variety of techniques \cite{wampler2010character, won2021control} create animations of competing agents (e.g. fight scenes) by computing optimal actions via planning algorithms (analogous to ``na\"ive planning'' in this paper). This often leads to believable behavior, but little to no artistic control.
Others \cite{shum2010simulating, funge1999cognitive, kapadia2016canvas} add various means for artists to guide the animation towards a desired state. These systems provide low-level artistic control over the space of outcomes, but still no high-level control over the animation's story arc. \citet{won2014generating} rank generated animations to show a human a diverse set of candidates to review.
In comparison, inverse inverse planning allows for a higher level of artistic control, automatically selecting the best result by modeling a human reviewer.

In interactive settings, the graphics community has long approached creating believable characters by requiring artists to manually give characters a large repertoire of high-level ``goals'' and low-level ``behaviors'' to express and accomplish those goals \cite{loyall1997believable, perlin1996improv, hayes1997acting, rousseau1998social, cassell2004beat}. In comparison, inverse inverse planning can \emph{automatically} derive ``behaviors'' to depict goals.

Some systems enforce structure on generated animations by searching over an ``evaluation function'' that proxies for aesthetic quality \cite{weyhrauch1997guiding, mateas1999oz, mateas2003faccade}. Such evaluation functions require significant low-level story-specific engineering effort for the artist to specify (dozens of pages of heuristics). In comparison, our method provides a general, principled tool for artists to write simple high-level ``evaluation functions.''

\paragraph{Textual story generation} As in graphics, the predominant approaches in textual story generation are planning- or logical-search-based \cite{meehan1977tale, lebowitz1985story, martens2013linear, riedl2010narrative}.
A variety of ad-hoc heuristics have emerged for modeling specific aspects of audience experience to guide planning.
For example, suspense can be measured by counting how few options a character has to solve a problem \cite{cheong2006computational, gerrig1994readers, bae2008use}, and characters can be made ``believable'' by ensuring that their actions' intents are visible to the reader and motivationally consistent \cite{riedl2004intent, szilas2003idtension}.
In relation to this line of work, inverse inverse planning is a general, principled framework that flexibly models a variety of audience inferences using Bayesian statistics. Audience modeling remains a challenge for newer language-model based approaches to textual story generation \cite{kreminski2022unmet}.

\section{Limitations and future work}
\label{sec:future-work}

\paragraph{Scalability}
The inverse inverse planner demonstrated in this paper is limited primarily by its scalability. The amount of computation needed grows with the size of the state space and number of characters (for ``planning''), the size of the audience's hypothesis space (for ``inverse planning''), and the complexity and length of scripts (for ``inverse inverse planning''). This is because, following \citeauthor{ullman2009help}, we used \emph{slow-but-exact} algorithms at every level for precision and robustness (e.g. value iteration for planning, enumerative inference for inverse planning). Nonetheless, all of the examples shown in this paper were generated within just a couple of minutes. This is because there were several opportunities for optimization: we precomputed the results of value iteration offline, and we parallelized beam search across many cores.

Having laid this groundwork with exact algorithms, we expect \emph{approximate} algorithms to help scale this framework to larger domains. As a first step, our ``miming'' domain demonstrates ways to scale inverse inverse planning to a large (indeed, continuous) state space: (1) approximating value functions with actor/critic neural networks, and (2) using gradient descent for the optimization step.

Approximate algorithms may additionally help scale inverse planning to larger \emph{hypothesis} spaces, allowing us to optimize over a richer space of audience inferences. For example, Bayesian inverse planning can be approximated using Sequential Monte Carlo (SMC) methods, which sample from the posterior distribution instead of exhaustively integrating over all hypotheses. \citet[Table 1(a)]{zhi2020online} show that compared to \citeauthor{ullman2009help}'s exact method (which we implement here), SMC runs 1-2 orders of magnitude faster across a benchmark of four different planning domains. Alternatively, the results of inverse planning can be approximated using amortized inference or deep learning. \citet[Table 5(b)]{malik2022social} train a spatiotemporal graph neural network to perform goal inference 4 orders of magnitude faster than the Bayesian inverse planner of \citet{netanyahu2021phase}. However, their method has roughly 10\% lower test accuracy than Bayesian inverse planning, suggesting that it might not generalize to match human intuitions well on the type of ``surprising'' out-of-distribution stories we seek here. What is the right balance between efficiency and accuracy? We leave further investigation in these directions to future work. A more efficient inverse inverse planner could enable not only scaling to larger problems, but also sophisticated real-time applications, for example in interactive fiction \citep[p.~153]{laurel1986toward} and human-computer improvisation \citep{pinhanez1999representation}.

\paragraph{Tools for artists}
In this paper, we showed how the language of optimization targets over Bayesian posteriors can be used  to express a desired audience experience. But of course, we do not expect artists will manually write these mathematical expressions---rather, we expect them to select and customize predefined story patterns (just as they do not manually program shaders/BRDFs, but rather select and customize predefined materials/textures).

This is possible because our optimization targets are abstract and portable across multiple domains. For example, we reuse the story arc pattern in both the kitchen and box-pulling domains (Sections \ref{sec:app-shapes} and \ref{sec:app-stumble}). Thus, there is much scope for future graphics/HCI work in designing intuitive interfaces for artists to select optimization targets. For example, an artist might select the story arc pattern from a library and then ``draw'' the desired arc on a tablet, or even describe it in natural language.

\paragraph{Modeling emotion}
Emotion is at the heart of storytelling. A promising future direction is to augment the audience model to reason about emotion: either to evoke certain emotions in the audience, or to use characters' visible emotions as degrees of freedom in story design. For example, if a villain captures a hero, showing the hero's sidekick smile covertly could make the audience infer that the sidekick was secretly aligned with the villain all along. In turn, this could evoke anger at the betrayal in the audience. Bayesian models analogous to inverse planning can capture some human emotional reasoning \cite{houlihan2022reasoning, saxe2017formalizing, ong2019computational, ong2015affective, ong2019applying}, and our ongoing work investigates methods for optimizing over these models.

\section{Conclusion}

We presented \emph{inverse inverse planning:} a principled computational framework for storytelling and animation, which is grounded in classic ideas from computer graphics, cognitive science, and literary theory, and goes beyond na\"ive simulation of rational agents. Building on an established model of social cognition (``inverse planning''), we showed how to optimize animations to evoke specific audience experiences (``inverse inverse planning''). We demonstrated the remarkable flexibility of inverse inverse planning by using it to capture a variety of storytelling elements, and presented experimental evidence that the resulting animations were effective depictions.

Our work lights the path to a first-principles approach to storytelling that treats audience experience as the ultimate desideratum. We do not rely on large datasets to learn storytelling techniques statistically, nor do we require ad-hoc manual encodings of those techniques. Rather, we show that those techniques emerge naturally as answers when we ask the right computational questions.

\begin{acks}
We thank the reviewers for their feedback, Google Cloud and the MIT subMIT team for offering generous computational resources, and Sam Cheyette, Karin Kukkonen, Sydney Levine, Alena Rote, Gabriella Safran, Rebecca Saxe, Tianmin Shu, Max Siegel, Andy Spielberg, and Tan Zhi-Xuan for thoughtful discussions regarding these ideas. This research was funded by NSF grants \#CCF-1231216, \#CCF-1723445 and \#2238839, ONR grant \#00010803, and supported by the Hertz Foundation, the Paul and Daisy Soros Fellowship, and an NSF Graduate Research Fellowship under grant \#1745302.
\end{acks}

\bibliographystyle{ACM-Reference-Format}
\bibliography{references,planning}


\begin{thebibliography}{66}


\ifx \showCODEN    \undefined \def \showCODEN     #1{\unskip}     \fi
\ifx \showDOI      \undefined \def \showDOI       #1{#1}\fi
\ifx \showISBNx    \undefined \def \showISBNx     #1{\unskip}     \fi
\ifx \showISBNxiii \undefined \def \showISBNxiii  #1{\unskip}     \fi
\ifx \showISSN     \undefined \def \showISSN      #1{\unskip}     \fi
\ifx \showLCCN     \undefined \def \showLCCN      #1{\unskip}     \fi
\ifx \shownote     \undefined \def \shownote      #1{#1}          \fi
\ifx \showarticletitle \undefined \def \showarticletitle #1{#1}   \fi
\ifx \showURL      \undefined \def \showURL       {\relax}        \fi
\providecommand\bibfield[2]{#2}
\providecommand\bibinfo[2]{#2}
\providecommand\natexlab[1]{#1}
\providecommand\showeprint[2][]{arXiv:#2}

\bibitem[Arora and Doshi(2021)]%
        {arora2021survey}
\bibfield{author}{\bibinfo{person}{Saurabh Arora} {and}
  \bibinfo{person}{Prashant Doshi}.} \bibinfo{year}{2021}\natexlab{}.
\newblock \showarticletitle{A survey of inverse reinforcement learning:
  Challenges, methods and progress}.
\newblock \bibinfo{journal}{\emph{Artificial Intelligence}}
  \bibinfo{volume}{297} (\bibinfo{year}{2021}), \bibinfo{pages}{103500}.
\newblock
\urldef\tempurl%
\url{https://arxiv.org/pdf/1806.06877}
\showURL{%
\tempurl}


\bibitem[Bae and Young(2008)]%
        {bae2008use}
\bibfield{author}{\bibinfo{person}{Byung-Chull Bae} {and}
  \bibinfo{person}{R~Michael Young}.} \bibinfo{year}{2008}\natexlab{}.
\newblock \showarticletitle{A use of flashback and foreshadowing for surprise
  arousal in narrative using a plan-based approach}. In
  \bibinfo{booktitle}{\emph{Joint international conference on interactive
  digital storytelling}}. Springer, \bibinfo{pages}{156--167}.
\newblock
\urldef\tempurl%
\url{https://www.academia.edu/download/30704103/icids1.pdf}
\showURL{%
\tempurl}


\bibitem[Baker et~al\mbox{.}(2008)]%
        {baker2008theory}
\bibfield{author}{\bibinfo{person}{Chris~L Baker}, \bibinfo{person}{Noah~D
  Goodman}, {and} \bibinfo{person}{Joshua~B Tenenbaum}.}
  \bibinfo{year}{2008}\natexlab{}.
\newblock \showarticletitle{Theory-based social goal inference}. In
  \bibinfo{booktitle}{\emph{Proceedings of the thirtieth annual conference of
  the cognitive science society}}. Citeseer, \bibinfo{pages}{1447--1452}.
\newblock
\urldef\tempurl%
\url{http://cocolab.stanford.edu/papers/BakerEtAl2008-Cogsci.pdf}
\showURL{%
\tempurl}


\bibitem[Baker et~al\mbox{.}(2009)]%
        {baker2009action}
\bibfield{author}{\bibinfo{person}{Chris~L Baker}, \bibinfo{person}{Rebecca
  Saxe}, {and} \bibinfo{person}{Joshua~B Tenenbaum}.}
  \bibinfo{year}{2009}\natexlab{}.
\newblock \showarticletitle{Action understanding as inverse planning}.
\newblock \bibinfo{journal}{\emph{Cognition}} \bibinfo{volume}{113},
  \bibinfo{number}{3} (\bibinfo{year}{2009}), \bibinfo{pages}{329--349}.
\newblock
\urldef\tempurl%
\url{https://www.sciencedirect.com/science/article/pii/S0010027709001607}
\showURL{%
\tempurl}


\bibitem[Baker et~al\mbox{.}(2007)]%
        {baker2007goal}
\bibfield{author}{\bibinfo{person}{Chris~L Baker}, \bibinfo{person}{Joshua~B
  Tenenbaum}, {and} \bibinfo{person}{Rebecca~R Saxe}.}
  \bibinfo{year}{2007}\natexlab{}.
\newblock \showarticletitle{Goal inference as inverse planning}. In
  \bibinfo{booktitle}{\emph{Proceedings of the Annual Meeting of the Cognitive
  Science Society}}, Vol.~\bibinfo{volume}{29}.
\newblock
\urldef\tempurl%
\url{https://escholarship.org/content/qt5v06n97q/qt5v06n97q.pdf}
\showURL{%
\tempurl}


\bibitem[Battaglia et~al\mbox{.}(2013)]%
        {battaglia2013simulation}
\bibfield{author}{\bibinfo{person}{Peter~W Battaglia},
  \bibinfo{person}{Jessica~B Hamrick}, {and} \bibinfo{person}{Joshua~B
  Tenenbaum}.} \bibinfo{year}{2013}\natexlab{}.
\newblock \showarticletitle{Simulation as an engine of physical scene
  understanding}.
\newblock \bibinfo{journal}{\emph{Proceedings of the National Academy of
  Sciences}} \bibinfo{volume}{110}, \bibinfo{number}{45}
  (\bibinfo{year}{2013}), \bibinfo{pages}{18327--18332}.
\newblock
\urldef\tempurl%
\url{https://www.pnas.org/doi/full/10.1073/pnas.1306572110}
\showURL{%
\tempurl}


\bibitem[Bellman(1966)]%
        {bellman1966dynamic}
\bibfield{author}{\bibinfo{person}{Richard Bellman}.}
  \bibinfo{year}{1966}\natexlab{}.
\newblock \showarticletitle{Dynamic programming}.
\newblock \bibinfo{journal}{\emph{Science}} \bibinfo{volume}{153},
  \bibinfo{number}{3731} (\bibinfo{year}{1966}), \bibinfo{pages}{34--37}.
\newblock


\bibitem[Cassell et~al\mbox{.}(2001)]%
        {cassell2004beat}
\bibfield{author}{\bibinfo{person}{Justine Cassell},
  \bibinfo{person}{Hannes~H\"{o}gni Vilhj\'{a}lmsson}, {and}
  \bibinfo{person}{Timothy Bickmore}.} \bibinfo{year}{2001}\natexlab{}.
\newblock \showarticletitle{BEAT: The Behavior Expression Animation Toolkit}.
\newblock In \bibinfo{booktitle}{\emph{Proceedings of the 28th Annual
  Conference on Computer Graphics and Interactive Techniques}}.
  \bibinfo{publisher}{Association for Computing Machinery},
  \bibinfo{address}{New York, NY, USA}, \bibinfo{pages}{477--486}.
\newblock
\showISBNx{158113374X}
\urldef\tempurl%
\url{https://doi.org/10.1145/383259.383315}
\showDOI{\tempurl}


\bibitem[Chandra et~al\mbox{.}(2022)]%
        {chandra2022designing}
\bibfield{author}{\bibinfo{person}{Kartik Chandra}, \bibinfo{person}{Tzu-Mao
  Li}, \bibinfo{person}{Joshua Tenenbaum}, {and} \bibinfo{person}{Jonathan
  Ragan-Kelley}.} \bibinfo{year}{2022}\natexlab{}.
\newblock \showarticletitle{Designing Perceptual Puzzles by Differentiating
  Probabilistic Programs}. In \bibinfo{booktitle}{\emph{Special Interest Group
  on Computer Graphics and Interactive Techniques Conference Proceedings
  (SIGGRAPH '22 Conference Proceedings)}}.
\newblock
\urldef\tempurl%
\url{https://doi.org/10.1145/3528233.3530715}
\showDOI{\tempurl}


\bibitem[Cheong and Young(2006)]%
        {cheong2006computational}
\bibfield{author}{\bibinfo{person}{Yun-Gyung Cheong} {and}
  \bibinfo{person}{R~Michael Young}.} \bibinfo{year}{2006}\natexlab{}.
\newblock \showarticletitle{A Computational Model of Narrative Generation for
  Suspense.}. In \bibinfo{booktitle}{\emph{AAAI}}. \bibinfo{pages}{1906--1907}.
\newblock
\urldef\tempurl%
\url{https://www.aaai.org/Papers/Workshops/2006/WS-06-04/WS06-04-003.pdf}
\showURL{%
\tempurl}


\bibitem[Dragan(2015)]%
        {dragan2015legible}
\bibfield{author}{\bibinfo{person}{Anca~D Dragan}.}
  \bibinfo{year}{2015}\natexlab{}.
\newblock \emph{\bibinfo{title}{Legible robot motion planning}}.
\newblock \bibinfo{thesistype}{Ph.\,D. Dissertation}. \bibinfo{school}{Carnegie
  Mellon University}.
\newblock
\urldef\tempurl%
\url{https://www.ri.cmu.edu/pub_files/2015/6/A_Dragan_Robotics_2015.pdf}
\showURL{%
\tempurl}


\bibitem[Durand et~al\mbox{.}(2002)]%
        {durand2002perceptual}
\bibfield{author}{\bibinfo{person}{Fr{\'e}do Durand}, \bibinfo{person}{Maneesh
  Agrawala}, \bibinfo{person}{Bruce Gooch}, \bibinfo{person}{Victoria
  Interrante}, \bibinfo{person}{Victor Ostromoukhov}, {and}
  \bibinfo{person}{Denis Zorin}.} \bibinfo{year}{2002}\natexlab{}.
\newblock \showarticletitle{Perceptual and artistic principles for effective
  computer depiction}.
\newblock \bibinfo{journal}{\emph{SIGGRAPH 2002 Course\# 13 Notes}}
  (\bibinfo{year}{2002}).
\newblock
\urldef\tempurl%
\url{http://people.csail.mit.edu/fredo/SIG02_ArtScience/DepictionNotes2.pdf}
\showURL{%
\tempurl}


\bibitem[Funge et~al\mbox{.}(1999)]%
        {funge1999cognitive}
\bibfield{author}{\bibinfo{person}{John Funge}, \bibinfo{person}{Xiaoyuan Tu},
  {and} \bibinfo{person}{Demetri Terzopoulos}.}
  \bibinfo{year}{1999}\natexlab{}.
\newblock \showarticletitle{Cognitive modeling: Knowledge, reasoning and
  planning for intelligent characters}. In
  \bibinfo{booktitle}{\emph{Proceedings of the 26th annual conference on
  Computer graphics and interactive techniques}}. \bibinfo{pages}{29--38}.
\newblock
\urldef\tempurl%
\url{https://dl.acm.org/doi/pdf/10.1145/311535.311538}
\showURL{%
\tempurl}


\bibitem[Gerrig and Bernardo(1994)]%
        {gerrig1994readers}
\bibfield{author}{\bibinfo{person}{Richard~J Gerrig} {and}
  \bibinfo{person}{Allan~BI Bernardo}.} \bibinfo{year}{1994}\natexlab{}.
\newblock \showarticletitle{Readers as problem-solvers in the experience of
  suspense}.
\newblock \bibinfo{journal}{\emph{Poetics}} \bibinfo{volume}{22},
  \bibinfo{number}{6} (\bibinfo{year}{1994}), \bibinfo{pages}{459--472}.
\newblock
\urldef\tempurl%
\url{https://www.sciencedirect.com/science/article/pii/0304422X94900213/pdf?md5=4e88a66f151c18e279f030fa56cba285&pid=1-s2.0-0304422X94900213-main.pdf}
\showURL{%
\tempurl}


\bibitem[Goodman and Frank(2016)]%
        {goodman2016pragmatic}
\bibfield{author}{\bibinfo{person}{Noah~D Goodman} {and}
  \bibinfo{person}{Michael~C Frank}.} \bibinfo{year}{2016}\natexlab{}.
\newblock \showarticletitle{Pragmatic language interpretation as probabilistic
  inference}.
\newblock \bibinfo{journal}{\emph{Trends in cognitive sciences}}
  \bibinfo{volume}{20}, \bibinfo{number}{11} (\bibinfo{year}{2016}),
  \bibinfo{pages}{818--829}.
\newblock
\urldef\tempurl%
\url{http://cocolab.stanford.edu/papers/GoodmanFrank2016-TICS.pdf}
\showURL{%
\tempurl}


\bibitem[Hayes-Roth et~al\mbox{.}(1997)]%
        {hayes1997acting}
\bibfield{author}{\bibinfo{person}{Barbara Hayes-Roth},
  \bibinfo{person}{Robert~van Gent}, {and} \bibinfo{person}{Daniel Huber}.}
  \bibinfo{year}{1997}\natexlab{}.
\newblock \showarticletitle{Acting in character}.
\newblock \bibinfo{journal}{\emph{Creating personalities for synthetic actors}}
  (\bibinfo{year}{1997}), \bibinfo{pages}{92--112}.
\newblock
\urldef\tempurl%
\url{https://citeseerx.ist.psu.edu/viewdoc/summary?doi=10.1.1.84.5760}
\showURL{%
\tempurl}


\bibitem[Heider and Simmel(1944)]%
        {heider1944experimental}
\bibfield{author}{\bibinfo{person}{Fritz Heider} {and}
  \bibinfo{person}{Marianne Simmel}.} \bibinfo{year}{1944}\natexlab{}.
\newblock \showarticletitle{An experimental study of apparent behavior}.
\newblock \bibinfo{journal}{\emph{The American journal of psychology}}
  \bibinfo{volume}{57}, \bibinfo{number}{2} (\bibinfo{year}{1944}),
  \bibinfo{pages}{243--259}.
\newblock
\urldef\tempurl%
\url{https://www.jstor.org/stable/pdf/1416950.pdf}
\showURL{%
\tempurl}


\bibitem[Ho et~al\mbox{.}(2021)]%
        {ho2021communication}
\bibfield{author}{\bibinfo{person}{Mark~K Ho}, \bibinfo{person}{Fiery Cushman},
  \bibinfo{person}{Michael~L Littman}, {and} \bibinfo{person}{Joseph~L
  Austerweil}.} \bibinfo{year}{2021}\natexlab{}.
\newblock \showarticletitle{Communication in action: Planning and interpreting
  communicative demonstrations.}
\newblock \bibinfo{journal}{\emph{Journal of Experimental Psychology: General}}
  (\bibinfo{year}{2021}).
\newblock
\urldef\tempurl%
\url{https://psyarxiv.com/a8sxk/}
\showURL{%
\tempurl}


\bibitem[Ho et~al\mbox{.}(2022)]%
        {ho2022planning}
\bibfield{author}{\bibinfo{person}{Mark~K Ho}, \bibinfo{person}{Rebecca Saxe},
  {and} \bibinfo{person}{Fiery Cushman}.} \bibinfo{year}{2022}\natexlab{}.
\newblock \showarticletitle{Planning with theory of mind}.
\newblock \bibinfo{journal}{\emph{Trends in Cognitive Sciences}}
  (\bibinfo{year}{2022}).
\newblock
\urldef\tempurl%
\url{https://saxelab.mit.edu/sites/default/files/publications/HoSaxeCushman2022.pdf}
\showURL{%
\tempurl}


\bibitem[Houlihan et~al\mbox{.}(2022)]%
        {houlihan2022reasoning}
\bibfield{author}{\bibinfo{person}{Sean~Dae Houlihan}, \bibinfo{person}{Desmond
  Ong}, \bibinfo{person}{Maddie Cusimano}, {and} \bibinfo{person}{Rebecca
  Saxe}.} \bibinfo{year}{2022}\natexlab{}.
\newblock \showarticletitle{Reasoning about the antecedents of emotions:
  Bayesian causal inference over an intuitive theory of mind}. In
  \bibinfo{booktitle}{\emph{Proceedings of the Annual Meeting of the Cognitive
  Science Society}}, Vol.~\bibinfo{volume}{44}.
\newblock
\urldef\tempurl%
\url{https://escholarship.org/content/qt7sn3w3n2/qt7sn3w3n2.pdf}
\showURL{%
\tempurl}


\bibitem[Jankovics(1974)]%
        {jankovics1974sisyphus}
\bibfield{author}{\bibinfo{person}{Marcell Jankovics}.}
  \bibinfo{year}{1974}\natexlab{}.
\newblock \bibinfo{title}{Sisyphus}.
\newblock
\newblock
\urldef\tempurl%
\url{https://www.youtube.com/watch?v=vyZK8rkeqPM}
\showURL{%
\tempurl}


\bibitem[Kapadia et~al\mbox{.}(2016)]%
        {kapadia2016canvas}
\bibfield{author}{\bibinfo{person}{Mubbasir Kapadia}, \bibinfo{person}{Seth
  Frey}, \bibinfo{person}{Alexander Shoulson}, \bibinfo{person}{Robert~W
  Sumner}, {and} \bibinfo{person}{Markus~H Gross}.}
  \bibinfo{year}{2016}\natexlab{}.
\newblock \showarticletitle{CANVAS: computer-assisted narrative animation
  synthesis.}. In \bibinfo{booktitle}{\emph{Symposium on Computer Animation}}.
  \bibinfo{pages}{199--209}.
\newblock
\urldef\tempurl%
\url{https://people.cs.rutgers.edu/~mk1353/pdfs/2016-sca-canvas.pdf}
\showURL{%
\tempurl}


\bibitem[Kiley et~al\mbox{.}(2016)]%
        {kiley2016game}
\bibfield{author}{\bibinfo{person}{Dilan~Patrick Kiley},
  \bibinfo{person}{Andrew~J Reagan}, \bibinfo{person}{Lewis Mitchell},
  \bibinfo{person}{Christopher~M Danforth}, {and}
  \bibinfo{person}{Peter~Sheridan Dodds}.} \bibinfo{year}{2016}\natexlab{}.
\newblock \showarticletitle{Game story space of professional sports: Australian
  rules football}.
\newblock \bibinfo{journal}{\emph{Physical Review E}} \bibinfo{volume}{93},
  \bibinfo{number}{5} (\bibinfo{year}{2016}), \bibinfo{pages}{052314}.
\newblock
\urldef\tempurl%
\url{https://cdanfort.w3.uvm.edu/research/2016-kiley-pre.pdf}
\showURL{%
\tempurl}


\bibitem[Kochenderfer et~al\mbox{.}(2022)]%
        {kochenderfer2022algorithms}
\bibfield{author}{\bibinfo{person}{Mykel~J Kochenderfer},
  \bibinfo{person}{Tim~A Wheeler}, {and} \bibinfo{person}{Kyle~H Wray}.}
  \bibinfo{year}{2022}\natexlab{}.
\newblock \bibinfo{booktitle}{\emph{Algorithms for decision making}}.
\newblock \bibinfo{publisher}{MIT press}.
\newblock
\urldef\tempurl%
\url{https://algorithmsbook.com}
\showURL{%
\tempurl}


\bibitem[Kreminski and Martens(2022)]%
        {kreminski2022unmet}
\bibfield{author}{\bibinfo{person}{Max Kreminski} {and} \bibinfo{person}{Chris
  Martens}.} \bibinfo{year}{2022}\natexlab{}.
\newblock \showarticletitle{Unmet Creativity Support Needs in Computationally
  Supported Creative Writing}. In \bibinfo{booktitle}{\emph{Proceedings of the
  First Workshop on Intelligent and Interactive Writing Assistants (In2Writing
  2022)}}.
\newblock
\urldef\tempurl%
\url{https://doi.org/10.18653/v1/2022.in2writing-1.11}
\showDOI{\tempurl}


\bibitem[Kukkonen(2014)]%
        {kukkonen2014bayesian}
\bibfield{author}{\bibinfo{person}{Karin Kukkonen}.}
  \bibinfo{year}{2014}\natexlab{}.
\newblock \showarticletitle{Bayesian narrative: Probability, plot and the shape
  of the fictional world}.
\newblock \bibinfo{journal}{\emph{Anglia}} \bibinfo{volume}{132},
  \bibinfo{number}{4} (\bibinfo{year}{2014}), \bibinfo{pages}{720--739}.
\newblock
\urldef\tempurl%
\url{https://www.duo.uio.no/bitstream/handle/10852/54629/1/01%2BKukkonen%2BBayesian%2BNarrative.pdf}
\showURL{%
\tempurl}


\bibitem[Laurel(1986)]%
        {laurel1986toward}
\bibfield{author}{\bibinfo{person}{Brenda Laurel}.}
  \bibinfo{year}{1986}\natexlab{}.
\newblock \emph{\bibinfo{title}{Toward the design of a computer-based
  interactive fantasy system}}.
\newblock \bibinfo{thesistype}{Ph.\,D. Dissertation}. \bibinfo{school}{The Ohio
  State University}.
\newblock
\urldef\tempurl%
\url{https://etd.ohiolink.edu/apexprod/rws_etd/send_file/send?accession=osu1487265143146814&disposition=inline}
\showURL{%
\tempurl}


\bibitem[Lebowitz(1985)]%
        {lebowitz1985story}
\bibfield{author}{\bibinfo{person}{Michael Lebowitz}.}
  \bibinfo{year}{1985}\natexlab{}.
\newblock \showarticletitle{Story-telling as planning and learning}.
\newblock \bibinfo{journal}{\emph{Poetics}} \bibinfo{volume}{14},
  \bibinfo{number}{6} (\bibinfo{year}{1985}), \bibinfo{pages}{483--502}.
\newblock
\urldef\tempurl%
\url{https://academiccommons.columbia.edu/doi/10.7916/D8K362MH/download}
\showURL{%
\tempurl}


\bibitem[Loyall(1997)]%
        {loyall1997believable}
\bibfield{author}{\bibinfo{person}{Aaron~B Loyall}.}
  \bibinfo{year}{1997}\natexlab{}.
\newblock \bibinfo{booktitle}{\emph{Believable Agents: Building Interactive
  Personalities.}}
\newblock \bibinfo{type}{{T}echnical {R}eport}. \bibinfo{institution}{Carnegie
  Mellon University, Department of Computer Science}.
\newblock
\urldef\tempurl%
\url{https://apps.dtic.mil/sti/pdfs/ADA327862.pdf}
\showURL{%
\tempurl}


\bibitem[Malik and Isik(2022)]%
        {malik2022social}
\bibfield{author}{\bibinfo{person}{Manasi Malik} {and} \bibinfo{person}{Leyla
  Isik}.} \bibinfo{year}{2022}\natexlab{}.
\newblock \showarticletitle{Social Inference from Relational Visual
  Information}.
\newblock \bibinfo{journal}{\emph{Journal of Vision}} \bibinfo{volume}{22},
  \bibinfo{number}{14} (\bibinfo{year}{2022}), \bibinfo{pages}{3810--3810}.
\newblock
\urldef\tempurl%
\url{https://jov.arvojournals.org/article.aspx?articleid=2784648}
\showURL{%
\tempurl}


\bibitem[Martens et~al\mbox{.}(2013)]%
        {martens2013linear}
\bibfield{author}{\bibinfo{person}{Chris Martens}, \bibinfo{person}{Anne-Gwenn
  Bosser}, \bibinfo{person}{Joao~F Ferreira}, {and} \bibinfo{person}{Marc
  Cavazza}.} \bibinfo{year}{2013}\natexlab{}.
\newblock \showarticletitle{Linear logic programming for narrative generation}.
  In \bibinfo{booktitle}{\emph{International Conference on Logic Programming
  and Nonmonotonic Reasoning}}. Springer, \bibinfo{pages}{427--432}.
\newblock


\bibitem[Mateas(1999)]%
        {mateas1999oz}
\bibfield{author}{\bibinfo{person}{Michael Mateas}.}
  \bibinfo{year}{1999}\natexlab{}.
\newblock \showarticletitle{An {O}z-centric review of interactive drama and
  believable agents}.
\newblock In \bibinfo{booktitle}{\emph{Artificial intelligence today}}.
  \bibinfo{publisher}{Springer}, \bibinfo{pages}{297--328}.
\newblock
\urldef\tempurl%
\url{https://citeseerx.ist.psu.edu/viewdoc/download?doi=10.1.1.47.9259&rep=rep1&type=pdf}
\showURL{%
\tempurl}


\bibitem[Mateas and Stern(2003)]%
        {mateas2003faccade}
\bibfield{author}{\bibinfo{person}{Michael Mateas} {and}
  \bibinfo{person}{Andrew Stern}.} \bibinfo{year}{2003}\natexlab{}.
\newblock \showarticletitle{Fa{\c{c}}ade: An experiment in building a
  fully-realized interactive drama}. In \bibinfo{booktitle}{\emph{Game
  developers conference}}, Vol.~\bibinfo{volume}{2}. \bibinfo{pages}{4--8}.
\newblock
\urldef\tempurl%
\url{https://www.cc.gatech.edu/fac/Charles.Isbell/classes/reading/papers/MateasSternGDC03.pdf}
\showURL{%
\tempurl}


\bibitem[Meehan(1977)]%
        {meehan1977tale}
\bibfield{author}{\bibinfo{person}{James~R Meehan}.}
  \bibinfo{year}{1977}\natexlab{}.
\newblock \showarticletitle{TALE-SPIN, An Interactive Program that Writes
  Stories.}. In \bibinfo{booktitle}{\emph{Ijcai}}, Vol.~\bibinfo{volume}{77}.
  \bibinfo{pages}{91--98}.
\newblock
\urldef\tempurl%
\url{https://www.ijcai.org/Proceedings/77-1/Papers/013.pdf}
\showURL{%
\tempurl}


\bibitem[Netanyahu et~al\mbox{.}(2021)]%
        {netanyahu2021phase}
\bibfield{author}{\bibinfo{person}{Aviv Netanyahu}, \bibinfo{person}{Tianmin
  Shu}, \bibinfo{person}{Boris Katz}, \bibinfo{person}{Andrei Barbu}, {and}
  \bibinfo{person}{Joshua~B Tenenbaum}.} \bibinfo{year}{2021}\natexlab{}.
\newblock \showarticletitle{{PHASE}: PHysically-grounded Abstract Social Events
  for Machine Social Perception}. In \bibinfo{booktitle}{\emph{Proceedings of
  the AAAI Conference on Artificial Intelligence}}, Vol.~\bibinfo{volume}{35}.
  \bibinfo{pages}{845--853}.
\newblock
\urldef\tempurl%
\url{https://www.tshu.io/PHASE/PHASE.pdf}
\showURL{%
\tempurl}


\bibitem[Ng et~al\mbox{.}(2000)]%
        {ng2000algorithms}
\bibfield{author}{\bibinfo{person}{Andrew~Y Ng}, \bibinfo{person}{Stuart
  Russell}, {et~al\mbox{.}}} \bibinfo{year}{2000}\natexlab{}.
\newblock \showarticletitle{Algorithms for inverse reinforcement learning.}. In
  \bibinfo{booktitle}{\emph{ICML}}, Vol.~\bibinfo{volume}{1}.
  \bibinfo{pages}{2}.
\newblock
\urldef\tempurl%
\url{http://www.datascienceassn.org/sites/default/files/Algorithms%20for%20Inverse%20Reinforcement%20Learning.pdf}
\showURL{%
\tempurl}


\bibitem[Ong et~al\mbox{.}(2019a)]%
        {ong2019applying}
\bibfield{author}{\bibinfo{person}{Desmond~C Ong}, \bibinfo{person}{Harold
  Soh}, \bibinfo{person}{Jamil Zaki}, {and} \bibinfo{person}{Noah~D Goodman}.}
  \bibinfo{year}{2019}\natexlab{a}.
\newblock \showarticletitle{Applying probabilistic programming to affective
  computing}.
\newblock \bibinfo{journal}{\emph{IEEE Transactions on Affective Computing}}
  \bibinfo{volume}{12}, \bibinfo{number}{2} (\bibinfo{year}{2019}),
  \bibinfo{pages}{306--317}.
\newblock
\urldef\tempurl%
\url{https://arxiv.org/pdf/1903.06445.pdf}
\showURL{%
\tempurl}


\bibitem[Ong et~al\mbox{.}(2015)]%
        {ong2015affective}
\bibfield{author}{\bibinfo{person}{Desmond~C Ong}, \bibinfo{person}{Jamil
  Zaki}, {and} \bibinfo{person}{Noah~D Goodman}.}
  \bibinfo{year}{2015}\natexlab{}.
\newblock \showarticletitle{Affective cognition: Exploring lay theories of
  emotion}.
\newblock \bibinfo{journal}{\emph{Cognition}}  \bibinfo{volume}{143}
  (\bibinfo{year}{2015}), \bibinfo{pages}{141--162}.
\newblock
\urldef\tempurl%
\url{https://www.sciencedirect.com/sdfe/reader/pii/S0010027715300196/pdf}
\showURL{%
\tempurl}


\bibitem[Ong et~al\mbox{.}(2019b)]%
        {ong2019computational}
\bibfield{author}{\bibinfo{person}{Desmond~C Ong}, \bibinfo{person}{Jamil
  Zaki}, {and} \bibinfo{person}{Noah~D Goodman}.}
  \bibinfo{year}{2019}\natexlab{b}.
\newblock \showarticletitle{Computational models of emotion inference in theory
  of mind: A review and roadmap}.
\newblock \bibinfo{journal}{\emph{Topics in cognitive science}}
  \bibinfo{volume}{11}, \bibinfo{number}{2} (\bibinfo{year}{2019}),
  \bibinfo{pages}{338--357}.
\newblock
\urldef\tempurl%
\url{https://onlinelibrary.wiley.com/doi/pdf/10.1111/tops.12371}
\showURL{%
\tempurl}


\bibitem[Pattanayak et~al\mbox{.}(2022)]%
        {pattanayak2022inverse}
\bibfield{author}{\bibinfo{person}{Kunal Pattanayak}, \bibinfo{person}{Vikram
  Krishnamurthy}, {and} \bibinfo{person}{Christopher Berry}.}
  \bibinfo{year}{2022}\natexlab{}.
\newblock \showarticletitle{Inverse-Inverse Reinforcement Learning. How to Hide
  Strategy from an Adversarial Inverse Reinforcement Learner}.
\newblock \bibinfo{journal}{\emph{arXiv preprint arXiv:2205.10802}}
  (\bibinfo{year}{2022}).
\newblock
\urldef\tempurl%
\url{https://arxiv.org/pdf/2205.10802}
\showURL{%
\tempurl}


\bibitem[Perlin and Goldberg(1996)]%
        {perlin1996improv}
\bibfield{author}{\bibinfo{person}{Ken Perlin} {and} \bibinfo{person}{Athomas
  Goldberg}.} \bibinfo{year}{1996}\natexlab{}.
\newblock \showarticletitle{Improv: A System for Scripting Interactive Actors
  in Virtual Worlds}. In \bibinfo{booktitle}{\emph{Proceedings of the 23rd
  Annual Conference on Computer Graphics and Interactive Techniques}}
  \emph{(\bibinfo{series}{SIGGRAPH '96})}. \bibinfo{publisher}{Association for
  Computing Machinery}, \bibinfo{address}{New York, NY, USA},
  \bibinfo{pages}{205--216}.
\newblock
\showISBNx{0897917464}
\urldef\tempurl%
\url{https://doi.org/10.1145/237170.237258}
\showDOI{\tempurl}


\bibitem[Pesowski et~al\mbox{.}(2020)]%
        {pesowski2020children}
\bibfield{author}{\bibinfo{person}{Madison~L Pesowski},
  \bibinfo{person}{Alyssa~D Quy}, \bibinfo{person}{Michelle Lee}, {and}
  \bibinfo{person}{Adena Schachner}.} \bibinfo{year}{2020}\natexlab{}.
\newblock \showarticletitle{Children use inverse planning to detect social
  transmission in design of artifacts}. In
  \bibinfo{booktitle}{\emph{Proceedings of the Annual Conference of the
  Cognitive Science Society}}.
\newblock
\urldef\tempurl%
\url{https://cognitivesciencesociety.org/cogsci20/papers/0150/0150.pdf}
\showURL{%
\tempurl}


\bibitem[Pinhanez(1999)]%
        {pinhanez1999representation}
\bibfield{author}{\bibinfo{person}{Claudio~S Pinhanez}.}
  \bibinfo{year}{1999}\natexlab{}.
\newblock \emph{\bibinfo{title}{Representation and recognition of action in
  interactive spaces}}.
\newblock \bibinfo{thesistype}{Ph.\,D. Dissertation}.
  \bibinfo{school}{Massachusetts Institute of Technology}.
\newblock
\urldef\tempurl%
\url{https://dspace.mit.edu/handle/1721.1/62342}
\showURL{%
\tempurl}


\bibitem[Qian et~al\mbox{.}(2021)]%
        {qian2021modeling}
\bibfield{author}{\bibinfo{person}{Yingdong Qian}, \bibinfo{person}{Marta
  Kryven}, \bibinfo{person}{Tao Gao}, \bibinfo{person}{Hanbyul Joo}, {and}
  \bibinfo{person}{Josh Tenenbaum}.} \bibinfo{year}{2021}\natexlab{}.
\newblock \showarticletitle{Modeling human intention inference in continuous 3D
  domains by inverse planning and body kinematics}.
\newblock \bibinfo{journal}{\emph{arXiv e-prints}} (\bibinfo{year}{2021}),
  \bibinfo{pages}{arXiv--2112}.
\newblock
\urldef\tempurl%
\url{https://social-intelligence-human-ai.github.io/docs/camready_12.pdf}
\showURL{%
\tempurl}


\bibitem[Radkani et~al\mbox{.}(2022)]%
        {radkani2022modeling}
\bibfield{author}{\bibinfo{person}{Setayesh Radkani}, \bibinfo{person}{Josh
  Tenenbaum}, {and} \bibinfo{person}{Rebecca Saxe}.}
  \bibinfo{year}{2022}\natexlab{}.
\newblock \showarticletitle{Modeling punishment as a rational communicative
  social action}. In \bibinfo{booktitle}{\emph{Proceedings of the Annual
  Meeting of the Cognitive Science Society}}, Vol.~\bibinfo{volume}{44}.
\newblock
\urldef\tempurl%
\url{https://escholarship.org/content/qt47g8d89h/qt47g8d89h.pdf}
\showURL{%
\tempurl}


\bibitem[Ramachandran and Amir(2007)]%
        {ramachandran2007bayesian}
\bibfield{author}{\bibinfo{person}{Deepak Ramachandran} {and}
  \bibinfo{person}{Eyal Amir}.} \bibinfo{year}{2007}\natexlab{}.
\newblock \showarticletitle{Bayesian Inverse Reinforcement Learning.}. In
  \bibinfo{booktitle}{\emph{IJCAI}}, Vol.~\bibinfo{volume}{7}.
  \bibinfo{pages}{2586--2591}.
\newblock
\urldef\tempurl%
\url{https://www.aaai.org/Papers/IJCAI/2007/IJCAI07-416.pdf}
\showURL{%
\tempurl}


\bibitem[Reagan et~al\mbox{.}(2016)]%
        {reagan2016emotional}
\bibfield{author}{\bibinfo{person}{Andrew~J Reagan}, \bibinfo{person}{Lewis
  Mitchell}, \bibinfo{person}{Dilan Kiley}, \bibinfo{person}{Christopher~M
  Danforth}, {and} \bibinfo{person}{Peter~Sheridan Dodds}.}
  \bibinfo{year}{2016}\natexlab{}.
\newblock \showarticletitle{The emotional arcs of stories are dominated by six
  basic shapes}.
\newblock \bibinfo{journal}{\emph{EPJ Data Science}} \bibinfo{volume}{5},
  \bibinfo{number}{1} (\bibinfo{year}{2016}), \bibinfo{pages}{1--12}.
\newblock
\urldef\tempurl%
\url{https://epjdatascience.springeropen.com/articles/10.1140/epjds/s13688-016-0093-1}
\showURL{%
\tempurl}


\bibitem[Riedl and Young(2004)]%
        {riedl2004intent}
\bibfield{author}{\bibinfo{person}{Mark~Owen Riedl} {and}
  \bibinfo{person}{R~Michael Young}.} \bibinfo{year}{2004}\natexlab{}.
\newblock \showarticletitle{An intent-driven planner for multi-agent story
  generation}. In \bibinfo{booktitle}{\emph{Autonomous Agents and Multiagent
  Systems, International Joint Conference on}}, Vol.~\bibinfo{volume}{2}. IEEE
  Computer Society, \bibinfo{pages}{186--193}.
\newblock
\urldef\tempurl%
\url{https://www.academia.edu/download/30704104/025_riedlm_story.pdf}
\showURL{%
\tempurl}


\bibitem[Riedl and Young(2010)]%
        {riedl2010narrative}
\bibfield{author}{\bibinfo{person}{Mark~O Riedl} {and}
  \bibinfo{person}{Robert~Michael Young}.} \bibinfo{year}{2010}\natexlab{}.
\newblock \showarticletitle{Narrative planning: Balancing plot and character}.
\newblock \bibinfo{journal}{\emph{Journal of Artificial Intelligence Research}}
   \bibinfo{volume}{39} (\bibinfo{year}{2010}), \bibinfo{pages}{217--268}.
\newblock
\urldef\tempurl%
\url{https://www.jair.org/index.php/jair/article/download/10669/25501}
\showURL{%
\tempurl}


\bibitem[Rousseau and Hayes-Roth(1998)]%
        {rousseau1998social}
\bibfield{author}{\bibinfo{person}{Daniel Rousseau} {and}
  \bibinfo{person}{Barbara Hayes-Roth}.} \bibinfo{year}{1998}\natexlab{}.
\newblock \showarticletitle{A Social-Psychological Model for Synthetic Actors}.
  In \bibinfo{booktitle}{\emph{Proceedings of the Second International
  Conference on Autonomous Agents}} (Minneapolis, Minnesota, USA)
  \emph{(\bibinfo{series}{AGENTS '98})}. \bibinfo{publisher}{Association for
  Computing Machinery}, \bibinfo{address}{New York, NY, USA},
  \bibinfo{pages}{165--172}.
\newblock
\showISBNx{0897919831}
\urldef\tempurl%
\url{https://doi.org/10.1145/280765.280795}
\showDOI{\tempurl}


\bibitem[Saxe and Houlihan(2017)]%
        {saxe2017formalizing}
\bibfield{author}{\bibinfo{person}{Rebecca Saxe} {and}
  \bibinfo{person}{Sean~Dae Houlihan}.} \bibinfo{year}{2017}\natexlab{}.
\newblock \showarticletitle{Formalizing emotion concepts within a Bayesian
  model of theory of mind}.
\newblock \bibinfo{journal}{\emph{Current opinion in Psychology}}
  \bibinfo{volume}{17} (\bibinfo{year}{2017}), \bibinfo{pages}{15--21}.
\newblock
\urldef\tempurl%
\url{https://daeh.info/assets/pubs/saxe2017cop.pdf}
\showURL{%
\tempurl}


\bibitem[Shafto et~al\mbox{.}(2014)]%
        {shafto2014rational}
\bibfield{author}{\bibinfo{person}{Patrick Shafto}, \bibinfo{person}{Noah~D
  Goodman}, {and} \bibinfo{person}{Thomas~L Griffiths}.}
  \bibinfo{year}{2014}\natexlab{}.
\newblock \showarticletitle{A rational account of pedagogical reasoning:
  Teaching by, and learning from, examples}.
\newblock \bibinfo{journal}{\emph{Cognitive psychology}}  \bibinfo{volume}{71}
  (\bibinfo{year}{2014}), \bibinfo{pages}{55--89}.
\newblock
\urldef\tempurl%
\url{https://www.sciencedirect.com/science/article/pii/S0010028514000024}
\showURL{%
\tempurl}


\bibitem[Shum et~al\mbox{.}(2010)]%
        {shum2010simulating}
\bibfield{author}{\bibinfo{person}{Hubert~PH Shum}, \bibinfo{person}{Taku
  Komura}, {and} \bibinfo{person}{Shuntaro Yamazaki}.}
  \bibinfo{year}{2010}\natexlab{}.
\newblock \showarticletitle{Simulating multiple character interactions with
  collaborative and adversarial goals}.
\newblock \bibinfo{journal}{\emph{IEEE Transactions on Visualization and
  Computer Graphics}} \bibinfo{volume}{18}, \bibinfo{number}{5}
  (\bibinfo{year}{2010}), \bibinfo{pages}{741--752}.
\newblock
\urldef\tempurl%
\url{https://ieeexplore.ieee.org/document/5669299}
\showURL{%
\tempurl}


\bibitem[Szilas(2003)]%
        {szilas2003idtension}
\bibfield{author}{\bibinfo{person}{Nicolas Szilas}.}
  \bibinfo{year}{2003}\natexlab{}.
\newblock \showarticletitle{IDtension: a narrative engine for Interactive
  Drama}. In \bibinfo{booktitle}{\emph{Proceedings of the technologies for
  interactive digital storytelling and entertainment (TIDSE) conference}}.
\newblock


\bibitem[Tauber and Steyvers(2011)]%
        {tauber2011using}
\bibfield{author}{\bibinfo{person}{Sean Tauber} {and} \bibinfo{person}{Mark
  Steyvers}.} \bibinfo{year}{2011}\natexlab{}.
\newblock \showarticletitle{Using inverse planning and theory of mind for
  social goal inference}. In \bibinfo{booktitle}{\emph{Proceedings of the
  Annual Meeting of the Cognitive Science Society}}, Vol.~\bibinfo{volume}{33}.
\newblock
\urldef\tempurl%
\url{https://cogsci.mindmodeling.org/2011/papers/0585/paper0585.pdf}
\showURL{%
\tempurl}


\bibitem[Thomas et~al\mbox{.}(1995)]%
        {thomas1995illusion}
\bibfield{author}{\bibinfo{person}{Frank Thomas}, \bibinfo{person}{Ollie
  Johnston}, {and} \bibinfo{person}{Frank Thomas}.}
  \bibinfo{year}{1995}\natexlab{}.
\newblock \bibinfo{booktitle}{\emph{The illusion of life: Disney animation}}.
\newblock \bibinfo{publisher}{Hyperion New York}.
\newblock
\urldef\tempurl%
\url{https://archive.org/details/TheIllusionOfLifeDisneyAnimation/}
\showURL{%
\tempurl}


\bibitem[Ullman et~al\mbox{.}(2009)]%
        {ullman2009help}
\bibfield{author}{\bibinfo{person}{Tomer Ullman}, \bibinfo{person}{Chris
  Baker}, \bibinfo{person}{Owen Macindoe}, \bibinfo{person}{Owain Evans},
  \bibinfo{person}{Noah Goodman}, {and} \bibinfo{person}{Joshua Tenenbaum}.}
  \bibinfo{year}{2009}\natexlab{}.
\newblock \showarticletitle{Help or hinder: Bayesian models of social goal
  inference}.
\newblock \bibinfo{journal}{\emph{Advances in neural information processing
  systems}}  \bibinfo{volume}{22} (\bibinfo{year}{2009}).
\newblock
\urldef\tempurl%
\url{https://www.tomerullman.org/papers/nips2010.pdf}
\showURL{%
\tempurl}


\bibitem[Vonnegut(2005)]%
        {vonnegut2011man}
\bibfield{author}{\bibinfo{person}{Kurt Vonnegut}.}
  \bibinfo{year}{2005}\natexlab{}.
\newblock \bibinfo{booktitle}{\emph{A Man Without a Country}}.
\newblock \bibinfo{publisher}{Seven Stories Press}.
\newblock


\bibitem[Wampler et~al\mbox{.}(2010)]%
        {wampler2010character}
\bibfield{author}{\bibinfo{person}{Kevin Wampler}, \bibinfo{person}{Erik
  Andersen}, \bibinfo{person}{Evan Herbst}, \bibinfo{person}{Yongjoon Lee},
  {and} \bibinfo{person}{Zoran Popovi\'{c}}.} \bibinfo{year}{2010}\natexlab{}.
\newblock \showarticletitle{Character Animation in Two-Player Adversarial
  Games}.
\newblock \bibinfo{journal}{\emph{ACM Trans. Graph.}} \bibinfo{volume}{29},
  \bibinfo{number}{3}, Article \bibinfo{articleno}{26} (\bibinfo{date}{jul}
  \bibinfo{year}{2010}), \bibinfo{numpages}{13}~pages.
\newblock
\showISSN{0730-0301}
\urldef\tempurl%
\url{https://doi.org/10.1145/1805964.1805970}
\showDOI{\tempurl}


\bibitem[Weyhrauch(1997)]%
        {weyhrauch1997guiding}
\bibfield{author}{\bibinfo{person}{Peter Weyhrauch}.}
  \bibinfo{year}{1997}\natexlab{}.
\newblock \bibinfo{booktitle}{\emph{Guiding interactive drama}}.
\newblock \bibinfo{publisher}{Carnegie Mellon University Pittsburgh}.
\newblock
\urldef\tempurl%
\url{http://cs.engr.uky.edu/~sgware/reading/papers/weyhrauch1997guiding.pdf}
\showURL{%
\tempurl}


\bibitem[Witkin and Kass(1988)]%
        {witkin1988spacetime}
\bibfield{author}{\bibinfo{person}{Andrew Witkin} {and}
  \bibinfo{person}{Michael Kass}.} \bibinfo{year}{1988}\natexlab{}.
\newblock \showarticletitle{Spacetime constraints}.
\newblock \bibinfo{journal}{\emph{ACM Siggraph Computer Graphics}}
  \bibinfo{volume}{22}, \bibinfo{number}{4} (\bibinfo{year}{1988}),
  \bibinfo{pages}{159--168}.
\newblock
\urldef\tempurl%
\url{https://dl.acm.org/doi/pdf/10.1145/378456.378507}
\showURL{%
\tempurl}


\bibitem[Won et~al\mbox{.}(2021)]%
        {won2021control}
\bibfield{author}{\bibinfo{person}{Jungdam Won}, \bibinfo{person}{Deepak
  Gopinath}, {and} \bibinfo{person}{Jessica Hodgins}.}
  \bibinfo{year}{2021}\natexlab{}.
\newblock \showarticletitle{Control Strategies for Physically Simulated
  Characters Performing Two-Player Competitive Sports}.
\newblock \bibinfo{journal}{\emph{ACM Trans. Graph.}} \bibinfo{volume}{40},
  \bibinfo{number}{4}, Article \bibinfo{articleno}{146} (\bibinfo{date}{jul}
  \bibinfo{year}{2021}), \bibinfo{numpages}{11}~pages.
\newblock
\showISSN{0730-0301}
\urldef\tempurl%
\url{https://doi.org/10.1145/3450626.3459761}
\showDOI{\tempurl}


\bibitem[Won et~al\mbox{.}(2014)]%
        {won2014generating}
\bibfield{author}{\bibinfo{person}{Jungdam Won}, \bibinfo{person}{Kyungho Lee},
  \bibinfo{person}{Carol O'Sullivan}, \bibinfo{person}{Jessica~K. Hodgins},
  {and} \bibinfo{person}{Jehee Lee}.} \bibinfo{year}{2014}\natexlab{}.
\newblock \showarticletitle{Generating and Ranking Diverse Multi-Character
  Interactions}.
\newblock \bibinfo{journal}{\emph{ACM Trans. Graph.}} \bibinfo{volume}{33},
  \bibinfo{number}{6}, Article \bibinfo{articleno}{219} (\bibinfo{date}{nov}
  \bibinfo{year}{2014}), \bibinfo{numpages}{12}~pages.
\newblock
\showISSN{0730-0301}
\urldef\tempurl%
\url{https://doi.org/10.1145/2661229.2661271}
\showDOI{\tempurl}


\bibitem[Xu et~al\mbox{.}(2022)]%
        {xu2022accelerated}
\bibfield{author}{\bibinfo{person}{Jie Xu}, \bibinfo{person}{Viktor
  Makoviychuk}, \bibinfo{person}{Yashraj Narang}, \bibinfo{person}{Fabio
  Ramos}, \bibinfo{person}{Wojciech Matusik}, \bibinfo{person}{Animesh Garg},
  {and} \bibinfo{person}{Miles Macklin}.} \bibinfo{year}{2022}\natexlab{}.
\newblock \showarticletitle{Accelerated policy learning with parallel
  differentiable simulation}.
\newblock \bibinfo{journal}{\emph{ICLR}} (\bibinfo{year}{2022}).
\newblock
\urldef\tempurl%
\url{https://arxiv.org/pdf/2204.07137.pdf}
\showURL{%
\tempurl}


\bibitem[Yoon et~al\mbox{.}(2016)]%
        {yoon2016talking}
\bibfield{author}{\bibinfo{person}{Erica~J Yoon},
  \bibinfo{person}{Michael~Henry Tessler}, \bibinfo{person}{Noah~D Goodman},
  {and} \bibinfo{person}{Michael~C Frank}.} \bibinfo{year}{2016}\natexlab{}.
\newblock \showarticletitle{Talking with tact: Polite language as a balance
  between kindness and informativity}. In \bibinfo{booktitle}{\emph{Proceedings
  of the 38th annual conference of the cognitive science society}}. Cognitive
  Science Society, \bibinfo{pages}{2771--2776}.
\newblock
\urldef\tempurl%
\url{http://socsci-dev.ss.uci.edu/~lpearl/courses/readings/YoonEtAl2016_Politeness.pdf}
\showURL{%
\tempurl}


\bibitem[Zhi-Xuan et~al\mbox{.}(2020)]%
        {zhi2020online}
\bibfield{author}{\bibinfo{person}{Tan Zhi-Xuan}, \bibinfo{person}{Jordyn
  Mann}, \bibinfo{person}{Tom Silver}, \bibinfo{person}{Josh Tenenbaum}, {and}
  \bibinfo{person}{Vikash Mansinghka}.} \bibinfo{year}{2020}\natexlab{}.
\newblock \showarticletitle{Online Bayesian Goal Inference for Boundedly
  Rational Planning Agents}.
\newblock \bibinfo{journal}{\emph{Advances in Neural Information Processing
  Systems}}  \bibinfo{volume}{33} (\bibinfo{year}{2020}).
\newblock
\urldef\tempurl%
\url{https://arxiv.org/pdf/2006.07532.pdf}
\showURL{%
\tempurl}


\end{thebibliography}

\newcommand{\comicwidth}[0]{0.9}

\begin{figure*}[t]
\centering
\includegraphics[page=3,width=\comicwidth\linewidth,trim={0 4.63in 0 0},clip]{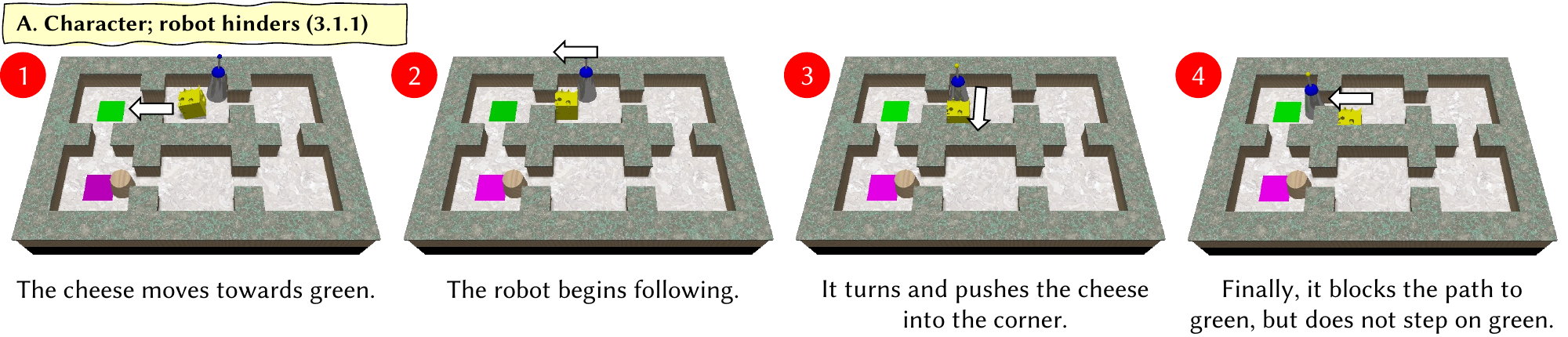}
\includegraphics[page=11,width=\comicwidth\linewidth,trim={0 4.63in 0 0},clip]{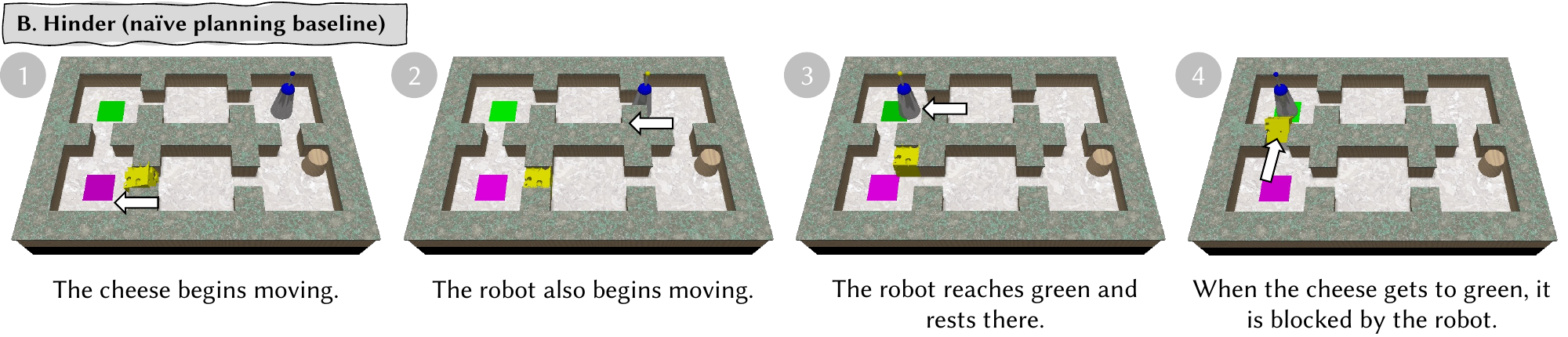}
\includegraphics[page=4,width=\comicwidth\linewidth,trim={0 4.63in 0 0},clip]{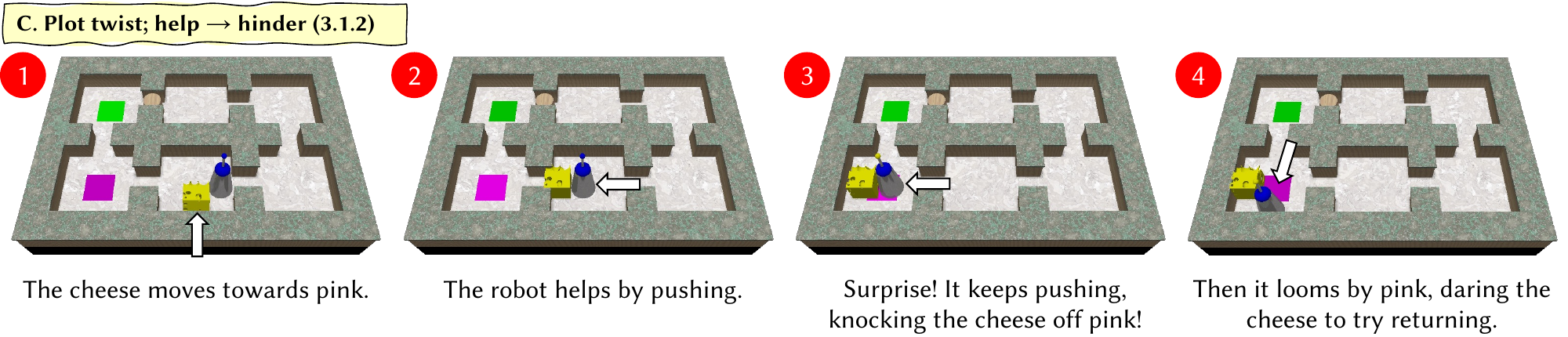}
\includegraphics[page=5,width=\comicwidth\linewidth,trim={0 4.63in 0 0},clip]{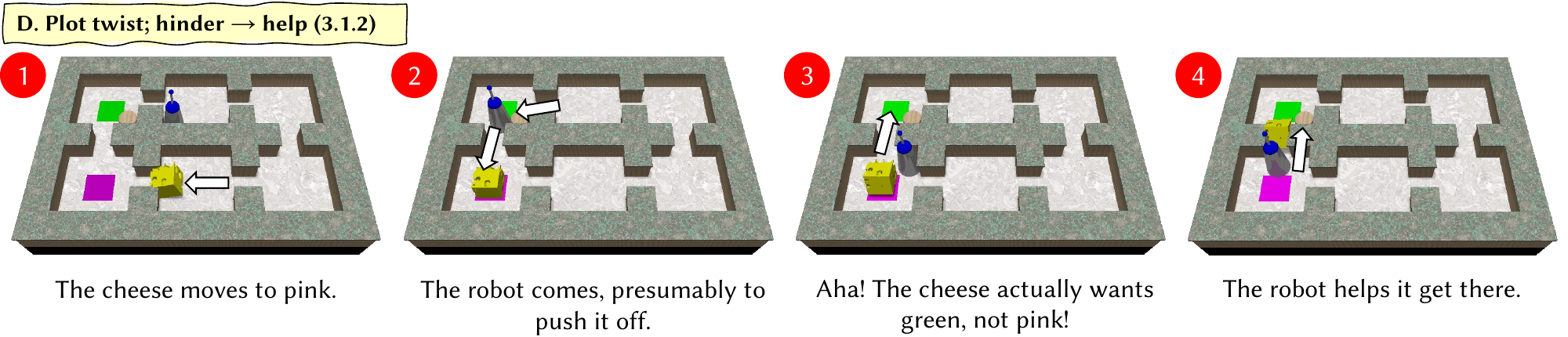}
\includegraphics[page=6,width=\comicwidth\linewidth,trim={0 4.63in 0 0},clip]{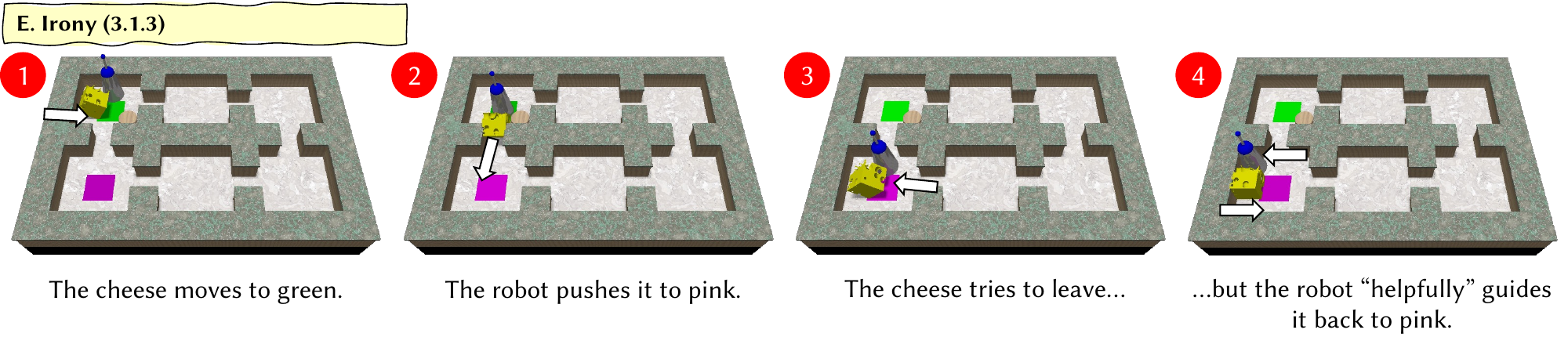}
\includegraphics[page=9,width=\comicwidth\linewidth,trim={0 4.63in 0 0},clip]{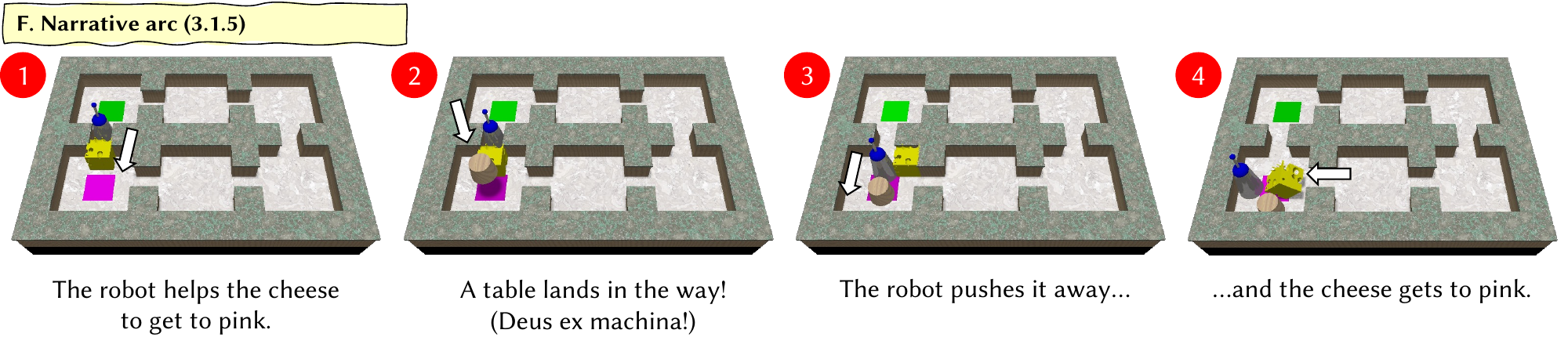}
\caption{Example animations generated in Section~\ref{sec:applications} (continued below). Grayed row shows the result of regular planning, for comparison.}\label{fig:example-animations-i}
\end{figure*}

\begin{figure*}[t]
\centering
\includegraphics[page=7,width=\comicwidth\linewidth,trim={0 4.63in 0 0},clip]{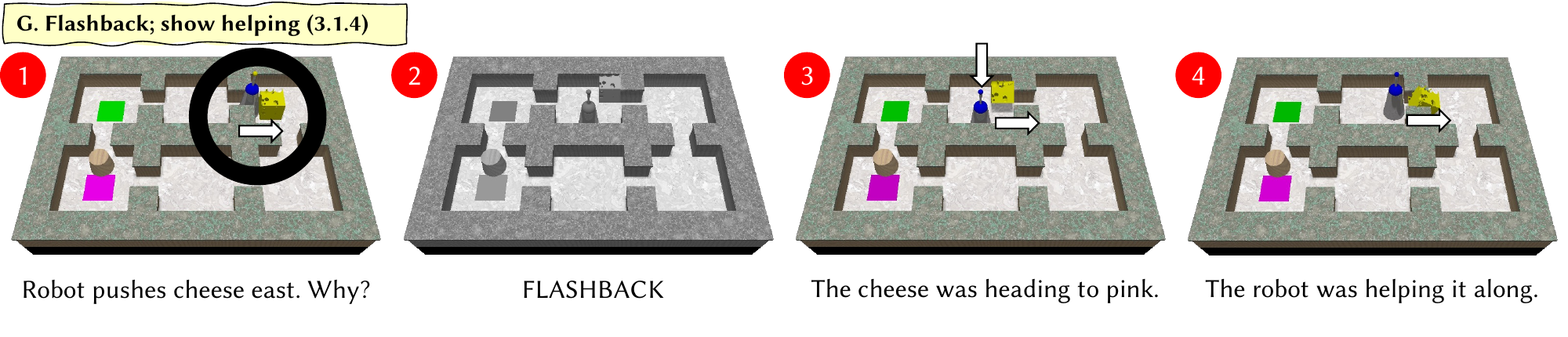}
\includegraphics[page=8,width=\comicwidth\linewidth,trim={0 4.63in 0 0},clip]{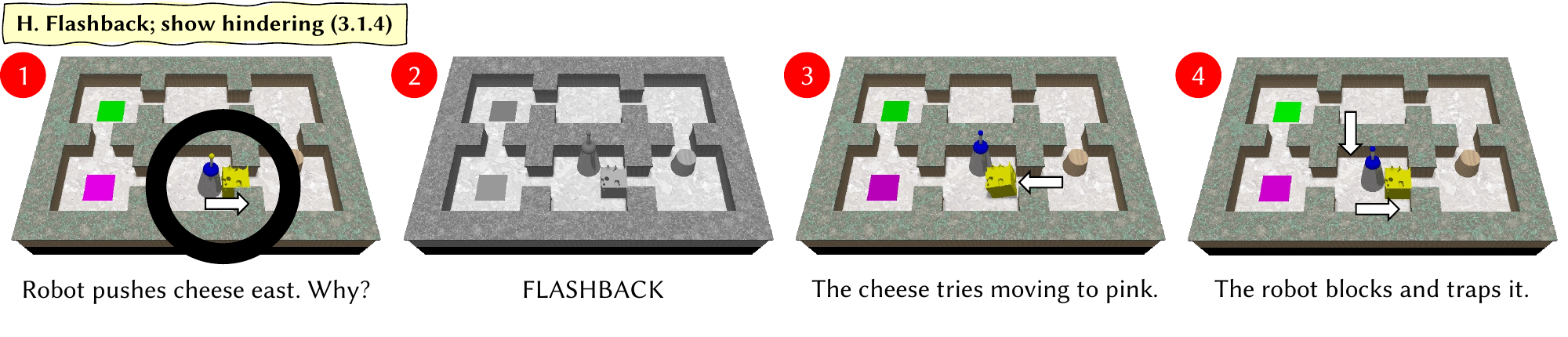}
\caption{Example animations generated in Section~\ref{sec:applications} (continued from above).}\label{fig:example-animations-ii}
\end{figure*}

\end{document}